\documentclass{aa} 

\usepackage{graphicx}
\usepackage{txfonts}
\usepackage{xcolor}
\usepackage{multirow}
\usepackage{array} 
\usepackage{array}[=2016-10-06]
\usepackage{siunitx}
\usepackage{booktabs}
\usepackage{longtable}
\usepackage{threeparttablex}

\usepackage{amsmath}
\usepackage{multirow} 
\usepackage{hyperref}
                                
\begin{document}

\newcommand{\vdag}{(v)^\dagger}

\def\xmm {\emph{XMM--Newton}}
\def\cxo {\emph{Chandra}}
\def\nustar {\emph{NuSTAR}}
\def\rst {\emph{ROSAT}}
\def\swift {\emph{Swift}}
\def\nicer {\emph{NICER}}
\def\hxmt {\emph{Insight}-HXMT}
\def\pks {Parkes}
\def\integral{\emph{INTEGRAL}}

\def\flux {\mbox{erg\,cm$^{-2}$\,s$^{-1}$}}
\def\lum {\mbox{erg\,s$^{-1}$}}
\def\nh {$N_{\rm H}$}
\def\kms  {\rm \ km \, s^{-1}}
\def\cms  {\rm \ cm \, s^{-1}}
\def\gs   {\rm \ g  \, s^{-1}}
\def\cmtre {\rm \ cm^{-3}}
\def\cmdue {\rm \ cm$^{-2}$}
\def\ss {\mbox{s\,s$^{-1}$}}
\def\chisq {$\chi ^{2}$}
\def\rchisq {$\chi_{r} ^{2}$}

\def\arc{\mbox{$^{\prime\prime}$}}
\def\arcmin{\mbox{$^{\prime}$}}
\def\deg{\mbox{$^{\circ}$}}

\def\rsun {~R_{\odot}}
\def\msun {~M_{\odot}}
\def\mdotav {\langle \dot {M}\rangle }

\def\uu {4U\,0142$+$614}
\def\ee {1E\,1048.1$-$5937}
\def\kes {1E\,1841$-$045}
\def\aa {1E\,1547$-$5408}
\def\axj {AX\,J1844$-$0258}
\def\rxs {1RXS\,J1708$-$4009}
\def\xte{XTE\,J1810$-$197}
\def\smc{CXOU\,J0100$-$7211\,}
\def\wessy{CXOU\,J1647$-$4552}
\def\ea {1E\,2259$+$586}
\def\ctb{CXOU\,J171405.7$-$381031}
\def\sgra{SGR\,1806$-$20}
\def\sgrb{SGR\,1900$+$14}
\def\sgrd{SGR\,1627$-$41}
\def\sgre{SGR\,0501$+$4516}
\def\sgrf{SGR\,1935+2154}
\def\lowba{SGR\,0418$+$5729}
\def\sgrg{SGR\,1833$-$0832}
\def\lowbb{Swift\,J1822.3$-$1606}
\def\galmag{SGR\,J1745$-$2900}
\def\sgras{Sgr\,A$^{\star}$}
\def\sgrh{SGR\,1801$-$21}
\def\sgri{SGR\,2013$+$34}
\def\psr{PSR\,1622$-$4950}
\def\hbpsr{PSR\,J1846$-$0258}
\def\radiohb{PSR\,J1119$-$6127}
\def\coronamag{Swift\,J1818.0$-$1607}
\def\sgrl{SGR\,J1830$-$0645}

\def\srcfirst {\mbox{Swift\,J1555.2$-$5402}}
\def\src {\mbox{Sw\,J1555}}

\title{A magnetar outburst with atypical evolution: the case of \srcfirst}

\author{
A.\ Borghese\inst{1}\fnmsep\thanks{ESA Research Fellow} 
\and 
F.\ Coti Zelati\inst{2,3,4} 
\and
M.\ Imbrogno\inst{2,3} 
\and
G.~L.\ Israel\inst{5} 
\and
D.\ De Grandis\inst{2,3}
\and 
D.~P.\ Pacholski\inst{6,7}
\and
M.\ Trudu\inst{8}
\and
M.\ Burgay\inst{8}
\and
S.\ Mereghetti\inst{6}
\and
N.\ Rea\inst{2,3}
\and
P.\ Esposito\inst{9,6}
\and
M.\ Pilia\inst{8}
\and
A.\ Possenti\inst{8}
\and
R.\ Turolla\inst{10,11}
\and
L.\ Ducci\inst{12}
}

\institute{
European Space Science (ESA), European Space Astronomy Center (ESAC), Camino Bajo del Castillo s/n, E-28692 Villanueva de la Ca\~{n}ada, Madrid
Spain
\email{alice.borghese@esa.int}
\and
Institute of Space Sciences (ICE, CSIC), Campus UAB, Carrer de Can Magrans s/n, E-08193, Barcelona, Spain
\and
Institut d'Estudis Espacials de Catalunya (IEEC), E-08860 Castelldefels (Barcelona), Spain
\and
INAF--Osservatorio Astronomico di Brera, Via Bianchi 46, Merate (LC), I-23807, Italy
\and
INAF--Osservatorio Astronomico di Roma, via Frascati 33, I-00078 Monteporzio Catone, Italy
\and
INAF--Istituto di Astrofisica Spaziale e Fisica Cosmica di Milano, via A.\ Corti 12, I-20133 Milano, Italy
\and
Universit\`a degli Studi di Milano Bicocca, Dipartimento di Fisica G. Occhialini, Piazza della Scienza 3, I-20126 Milano, Italy
\and
INAF--Osservatorio Astronomico di Cagliari, Via della Scienza 5, I-09047 Selargius, Italy
\and
Scuola Universitaria Superiore IUSS Pavia, Piazza della Vittoria 15, I-27100 Pavia, Italy
\and
Dipartimento di Fisica e Astronomia, Universit\`a degli Studi di Padova, via F. Marzolo 8, I-35131 Padova, Italy
\and
Mullard Space Science Laboratory, University College London, Holmbury St. Mary, Dorking, Surrey RH5 6NT, UK
\and
Institut f\"ur Astronomie und Astrophysik, Universit\"at T\"ubingen, Sand 1, D-72076 T\"ubingen, Germany
}

\date{Received October 8, 2025}

\abstract
{The magnetar \srcfirst\ was discovered in outburst on 2021 June 3 by the Burst Alert Telescope on board the \swift\ satellite. Early X‐ray follow‐up revealed a spin period $P\simeq3.86$\,s, a period derivative $\dot P\simeq3\times10^{-11}$\,s\,s$^{-1}$, dozens of short bursts, and an unusually flux decline. We report here on the X‐ray monitoring of \srcfirst\ over the first $\simeq$29 months of its outburst with \swift, \nicer, \nustar, \integral\ and \hxmt, as well as radio observations with \pks\ soon after the outburst onset. The observed 0.3–10 keV flux remained at levels $\gtrsim10^{-11}$ erg cm$^{-2}$ s$^{-1}$ for nearly 500 days before dropping by a factor of $\simeq 10$ from its June 2021 peak towards the end of the monitoring campaign. During this time span, the spectrum was dominated by a single blackbody, with temperature attaining approximately a constant value ($\sim$1.2\,keV) while the inferred radius shrank from $\approx1.7$\,km to $\approx0.3$\,km (assuming a source distance of 10\,kpc). The long‐term spin‐down rate ($\dot P\simeq3.6\times10^{-11}$\,s\,s$^{-1}$) is only $\sim15\%$ higher than that measured in the first 30 days. No periodic or burst‐like radio emission was detected, in line with what has been previously reported using different radio facilities. 
The persistently high temperature, shrinking hotspot, and a prolonged bright flux plateau followed by a fast dimming observed during the outburst evolution pose a challenge for the outburst mechanisms proposed so far.}


    \maketitle

\section{Introduction}
\label{sec:intro}

Magnetars are isolated X-ray pulsars with luminosity $L_X\sim10^{31}$--10$^{36}$\,\lum\ whose emission is believed to be powered by the dissipation of their magnetic energy (for reviews, see \citealt{turolla15,kaspi17,esposito21,rea25}). Their most spectacular observational manifestations are bursts of X-/gamma-rays and X-ray outbursts. The former are observed on timescales from milliseconds to minutes, reaching peak luminosities $L_X\sim10^{38}$--10$^{46}$\,\lum\ (e.g., \citealt{collazzi15}). The latter are longer-lived episodes where the persistent luminosity increases by factors of few to thousands up to $L_X\sim10^{34}$--10$^{36}$\,\lum\ and then usually decreases back to quiescence on time scales of months/years (see, e.g., the Magnetar Outburst Online Catalog\footnote{\url{http://magnetars.ice.csic.es}}; \citealt{cotizelati18}). The onset of an outburst is typically accompanied by and discovered through the emission of short bursts. The decay pattern differs from outburst to outburst, but it is often characterised by a rapid initial decay within hours to a few days, followed by a slower fading that can be modelled by a power-law or exponential functions. Despite the great diversity in decay profiles, all outbursts share some common trends. For instance, a higher luminosity at the outburst onset typically corresponds to a larger amount of energy released during the entire event (generally in the range $\sim10^{41}-10^{43}$\,erg), and outbursts with a longer overall duration tend to be the most energetic ones \citep{cotizelati18}. Outbursts are most likely caused by a sudden release of heat in a restricted area within or just above the magnetar crust, leading to the formation of a hot spot that progressively cools down. However, the exact triggering mechanism is still an open question: it may involve localised internal magnetic stresses that deform part of the crust \citep[see, e.g.,][]{dehman2020, degrandis22} or a twisted bundle in the magnetosphere \citep[see, e.g.,][]{beloborodov09, carrasco2019}.

\srcfirst\ (henceforth \src) was discovered on 2021 June 3 at 09:45:46 UT, when the Burst Alert Telescope (BAT) on board the \emph{Neil Gehrels Swift Observatory} (\swift) triggered and localized a short burst of X-rays similar to those emitted by magnetars \citep{palmer21a}. Over the following hours, an X-ray counterpart was discovered using the \swift\ X-ray Telescope (XRT) and a periodic modulation of its X-ray emission with $P\sim$3.86\,s was measured using the \emph{Neutron star Interior Composition Explorer} (\nicer). This confirmed the magnetar nature of the source \citep{cotizelati21}.
\nicer\ observed \src\ with a daily cadence over the first month of its outburst, revealing a thermal spectrum at a nearly constant flux of $\sim4\times10^{-11}$\,\flux\ in the soft X-ray band throughout this time span. Emission up to energies of $\sim$40\,keV with a spectral shape well described by a power law was also detected using the \emph{Nuclear Spectroscopic Telescope Array} (\nustar). The \nicer\ observations allowed for a measurement of the spin period derivative, $\dot{P} \simeq 3\times10^{-11}$\,\ss, and also revealed a number of bursts from the source (\citealt{enoto21b}; see also \citealt{bernardini21,klingler21,zhang21}).

This paper presents X-ray observations of \src\ using \nicer, \swift/XRT, \hxmt, \nustar\ and \integral\ over the first 29 months of the outburst as well as radio observations performed using \pks\ soon after the outburst onset (Sect.\,\ref{sec:obs}). We report on: (i) the X-ray spectral and timing properties of \src\ (Sect.\,\ref{sec:xrayanalysis}, \ref{sec:broadbandspec} and \ref{sec:timing}) as well as on the results of our searches for short X-ray bursts (Sect.\,\ref{sec:bursts}); (ii) searches for periodic and/or bursting radio emission in observations with \pks/Murriyang close to the peak of the outburst (Sect.\,\ref{sec:radio}). A discussion of the results and conclusions follow (Sect.\,\ref{sec:discussion}).

\section{Observations and data analysis} 
\label{sec:obs}
\subsection{X-ray observations} 

Table\,\ref{tab:obsX} reports the log of the X-ray observations used in this work. Additional \nicer\ and \nustar\ observations were carried out during the first month of the outburst, and were analysed by \cite{enoto21b}. 

Data reduction was performed with tools provided in {\sc HEASoft} (v.6.33) and {\sc HXMTdas} (v.2.06). 
Photon arrival times were barycentred using the \swift/XRT enhanced position, R.A. = 15$^\mathrm{h}$55$^\mathrm{m}$08$\fs$66, Decl. = --54$^{\circ}$03$^{\prime}$41$\farcs$1 (J2000.0; uncertainty of 2.2\arc\ at 90\% c.l.;  \citealt{evans21}) and the JPL planetary ephemeris DE430. The spectral analysis was performed using {\sc Xspec} \citep{arnaud96}, applying the {\sc Tbabs} model with cross-sections of \citet{verner96} and elemental abundances of \citet{wilms00} to describe the effects of interstellar absorption and the convolution model {\sc Cflux} to estimate the source flux. In the following, we quote all uncertainties at 1$\sigma$ confidence level (c.l.) and assume a distance to the source of 10\,kpc.

\subsubsection{Swift}\label{sec:swift}

The \swift/XRT (\citealt{burrows05}) monitored \src\ with 85 pointings between 2021 June 3 and 2023 October 21. Most observations were performed with the XRT in the windowed timing mode (WT; time resolution of 1.8\,ms) to study the time evolution of the spin signal. Five sparse observations between 2021 June 3 and 2022 January 5 and all observations since 2022 mid-August\footnote{The source had become too faint for the WT mode at these epochs for a meaningful spectral analysis.} were performed instead in photon counting mode (PC; 2.5\,s). 
We collected source photons within a circle with a radius of 20 pixels (1 pixel = 2$\farcs$36) and background counts from an annulus with radii of 80 and 120 pixels for data in the WT mode and 40 and 80 pixels for data in PC mode. For the spectral analysis, we selected events with grades 0--12 and 0 for PC and WT data, respectively, while we extended the timing analysis to events with grade 0--2 for WT data sets.

The \swift/XRT background-subtracted spectra were grouped to have at least 10 counts in each spectral channel up to observation ID: 00014971022. Due to the low photon counting statistics in the following observations, we opted for binning the remaining \swift\ spectra according to a variable minimum number of counts between 5 and 3 counts per spectral bin. The $W$-statistic was employed for model parameter estimation and error calculation for all \swift/XRT spectra. 

\subsubsection{NICER}\label{sec:nicer}
\nicer\ \citep{gendreau12} observed the field of \src\ for $\sim$2.5\,ks starting on 2021 June 3 at 11:20 UT, just $\sim$1.6\,h after the first \swift/BAT trigger. The source was subsequently monitored over a time interval of $\sim$4.5 months until 2021 October 22, just before entering a Solar constraint period. A dozen additional observations were performed between 2022 July 21 and 2022 August 17. In this work, we only include the data we obtained through our own Target of Opportunity requests (see Table\,\ref{tab:obsX} for a log of the observations).
We processed and screened the data using the {\sc nicerl2} tool. Then, we extracted background-subtracted spectra and light curves using the {\sc nicerl3-spect} and {\sc nicerl3-lc} tools, respectively, with the {\sc scorpeon} background model\footnote{\url{https://heasarc.gsfc.nasa.gov/docs/nicer/analysis_threads/scorpeon-overview/}}.
The spectra were binned to guarantee at least 100 background-subtracted counts per energy bin so as to apply the $\chi^2$ statistics.

\subsubsection{NuSTAR}\label{sec:nustar}
\src\ was observed by \nustar\ \citep{harrison13} at four epochs. We report here on the last observation performed on 2021 October 7 \citep[see][for details on the first three pointings]{enoto21b}. We applied standard analysis threads to reprocess the raw data for the two focal plane detectors, referred to as FPMA and FPMB.
We used the tool {\sc optimize\_radius\_snr} of the NuSTAR-gen-utils package 
\citep{brian_grefenstette_2025_14969199},   to estimate the source extraction radius that maximises the S/N ratio. This resulted in circular source extraction regions with radii of 79 arcsec for FPMA and 74 arcsec for FPMB.
Background counts were accumulated from a nearby source-free circular region  with  70 arcsec radius.
We then generated light curves, background-subtracted spectra and response files for both FPMs with the script {\sc nuproducts}. We rebinned the data with {\sc ftgrouppha} following the Kaastra \& Bleeker optimal binning algorithm \citep{kaastra16} in order to have at least 25 counts per bin. The source was detected up to $\sim$20\,keV at a net count rate of $\sim$1.34\,counts\,s$^{-1}$ (summing up the two FPMs).

\subsubsection{Insight-HXMT}\label{sec:hxmt}

The Hard X-ray Modulation Telescope (\hxmt; \citealt{zhang20}) observed \src\ for $\sim$100\,ks starting on 2021 June 10 at 17:38:32 UT, using the Low Energy telescope (LE), the Medium energy telescope (ME) and the High energy telescope (HE). We used the {\sc hpipeline} to process and screen the data, to create response files and to extract background-subtracted spectra and light curves. \src\ was detected at a very low signal-to-noise ratio in these data. Only LE data were used for the analysis, only to refine our timing solution (see Table\,\ref{tab:obsX} for the corresponding on-source exposure time and net count rate after data screening). 

\subsubsection{INTEGRAL}

The position of \src\ was extensively covered by observations with the Imager on Board the \integral\ Satellite \citep[IBIS;][]{ubertini03}.
We selected all the Science Windows\footnote{\integral\ observations are divided into science windows, i.e.  pointings with typical durations of $\sim$2--3\,ks.} (ScWs) in the public archive where the source was located within 14.5$^{\circ}$ from the centre of the IBIS field of view.  The selected ScWs were screened for high or variable background. The filtering resulted in a total of 11120 ScWs, corresponding to an observation time of 24.65\,Ms before the outburst (from March 2003 to 2021 June 3) and 1.96\,Ms divided into 806 ScWs since the outburst onset up to 27 August 2023.

During the first 150 days after the \swift/BAT trigger, \integral\ observed the source in the central field of view for 58.8\,ks. Using OSA \citep[v.11.2;][]{goldwurm03}, we created an image in the 30--80\,keV energy interval using all available observations during this period. The source was not detected; we estimate a 3$\sigma$ upper limit on its flux of $9.7\times10^{-11}$\,\flux.

\subsection{Radio observations}

Three visits were performed using the Ultra Wideband Low-frequency receiver (UWL; \citealt{hobbs20}) of the 64-m \pks\ Murriyang Radio telescope (Table\,\ref{tab:obsradio}). The data were collected with two backends in parallel. The {\it Medusa} backend \citep{hobbs20} was always operated over a 3.3\,GHz bandwidth centred at 2.4\,GHz, with a frequency resolution of 1\,MHz, a sampling time of 128\,\textmu s, and a 2-bit digitisation. On 2021 June 4 and 7, the PDFB4 backend acquired 256\,MHz of data centred at a frequency of 1369\,MHz, with a frequency resolution of 0.5\,MHz. On 2021 June 5, data were recorded over a 1024\,MHz bandwidth, centred at 3100\,MHz, with a frequency resolution of 2\,MHz. In all cases, the PDFB4 time series were 2-bit sampled every 256\,\textmu s. 

\section{Results}
\label{results}

\subsection{X-ray monitoring}
\label{sec:xrayanalysis}

\begin{figure*}
\centering
\includegraphics[width=1.9\columnwidth]{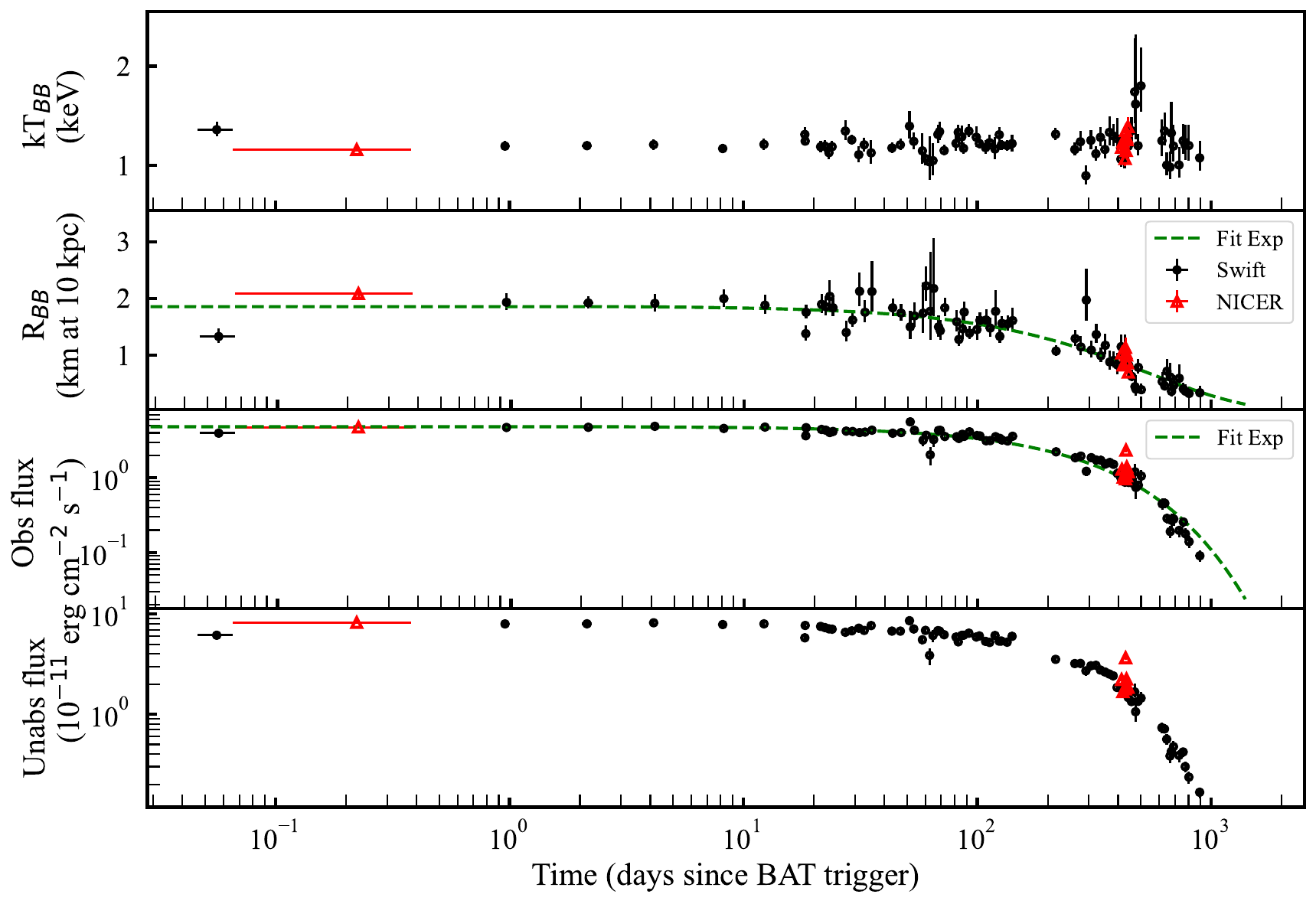} 
\caption{\label{fig:spec_evo}Temporal evolution of the blackbody temperature and radius, and of the observed and unabsorbed flux (0.3--10\,keV) of \srcfirst\ over a time span of about 900 days since the epoch of the first \swift/BAT trigger (on 2021 June 3 at 09:45:46 UT; 59368.40678\,MJD). The green dashed lines indicate the fit with an exponential function for the radius and observed flux.}
\end{figure*}

We fit an absorbed blackbody to the available \nicer\ and \swift/XRT spectra jointly, tying the hydrogen column density \nh\ among the different data sets and allowing the other parameters to vary. The fit gave an overall satisfactory description of the data with \nh = (8.4$\pm$0.2)$\times10^{22}$\,\cmdue (null hypothesis probability\footnote{The null hypothesis probability (nhp) represents the probability that the deviations between the data and the model are due to chance alone. In general, a model can be rejected when the nhp is smaller than 0.05.}=0.5).
Figure\,\ref{fig:spec_evo} shows the temporal evolution of the temperature and radius of the emitting region, and of the observed and unabsorbed flux in the 0.3--10\,keV energy interval. The blackbody temperature $kT_{\rm BB}$ did not display any significant variability during the whole duration of our monitoring campaign, attaining an average value of $\sim$1.2\,keV. The corresponding radius $R_{\rm BB}$ slowly decreased from $\sim$1.7\,km to $\sim$0.3\,km over $\sim$895\,days, and its temporal evolution can be approximated as an exponential function with an $e$-folding time of $\tau=538\pm21$\,days (chi-square \chisq =189 for 79 degrees of freedom (dof)). The observed 0.3--10\,keV flux gradually decreased from $\sim$5$\times$10$^{-11}$\,\flux\ to $\sim$1$\times$10$^{-11}$\,\flux\ during the first $\sim$500 days of our monitoring campaign. Then, a sharper decline followed, with the flux dropping to $\sim$9$\times$10$^{-13}$\,\flux\ on 2023 October 21 (the epoch of the last observation included in this campaign). The temporal evolution of the observed flux can be described by an exponential function with an $e$-folding time of $\tau=262\pm3$\,days. Although the fit does not yield a statistically acceptable result (\chisq =402 for 79 dof), this model adequately captures the general trend of the decay. We attribute the high \chisq\ obtained to the scatter in the data points and/or small-amplitude variability superimposed on the exponential decay. It is worthy to note that the decay time of the flux is in agreement with the characteristic time for the radius shrinking assuming blackbody emission at constant temperature.

The sky position of \src\ was serendipitously observed with \swift/XRT a few times before 2021 June. The source was not detected in any observations with a 3$\sigma$ upper limit on the averaged count rate of 0.007\,counts\,s$^{-1}$, derived from the stacking of the available archival pointings (see Table\,\ref{tab:obsX}). Using the {\sc webpimms} tool\footnote{\url{https://heasarc.gsfc.nasa.gov/cgi-bin/Tools/w3pimms/w3pimms.pl}} and assuming an absorbed blackbody with \nh=8.4$\times$10$^{22}$\,\cmdue\ and $kT=0.3$\,keV, the upper limit translates to a 0.3--10\,keV observed flux of $<2.7\times10^{-13}$\,\flux, which corresponds to a luminosity of $<10^{35}$\,\lum\ at 10\,kpc. 

We inspected the observations performed after a burst trigger for the presence of diffuse emission around the source, which may be associated with a rapidly evolving component causally connected with the burst emission, such as a dust scattering halo \citep[see e.g.,][]{tiengo10, mereghetti20}. Our searches did not yield significant detection.

\subsection{Broad-band spectrum}
\label{sec:broadbandspec}

We analysed the broadband spectrum of the source using the latest \nustar\ observation, jointly fitted with the \swift/XRT spectrum obtained simultaneously (OsbID: 00014352040). The spectral fitting was limited to energies below 20 keV, where the source count rate exceeds the background level. For the spectral analysis, we included a renormalisation factor to account for cross-calibration uncertainties, which was fixed at 1 for \nustar/FPMA and allowed to vary for \nustar/FPMB and \swift/XRT. A model composed of only an absorbed blackbody did not provide a good fit, revealing structured residuals above $\sim$10\,keV. The addition of a power law to the model significantly improved the fit to a reduced chi-square \rchisq = 210 for 213 dof. The absorption column density was frozen at \nh=$8.4\times10^{22}$\,cm$^{-2}$ (see Sec.\,\ref{sec:xrayanalysis}). The fit yielded the following values for the constant: $1.00\pm0.01$ for \nustar/FPMB and $0.92\pm0.04$ for \swift/XRT. Best-fitting parameters were: $kT_{\rm BB}$ = $1.16\pm0.01$\,keV, $R_{\rm BB}$ = $1.71^{+0.01}_{-0.02}$\,km and photon index $\Gamma$ = $0.89^{+0.50}_{-0.49}$. The parameters for the blackbody component are consistent with those derived from the fitting of the soft X-rays spectra extracted from adjacent observations. The $10-60$\,keV flux was $(7.19^{+0.41}_{-1.38})\times10^{-12}$\,\flux, indicating a $\sim20\%$ decrease compared to the previous \nustar\ observation performed three months earlier, which measured a flux of $(8.72\pm0.85)\times10^{-12}$\,\flux\ \citep{enoto21b}.

\subsection{Timing analysis and phase-resolved spectroscopy}
\label{sec:timing}

\begin{figure}[htbp!]
  \centering
\includegraphics[width=1.\columnwidth]{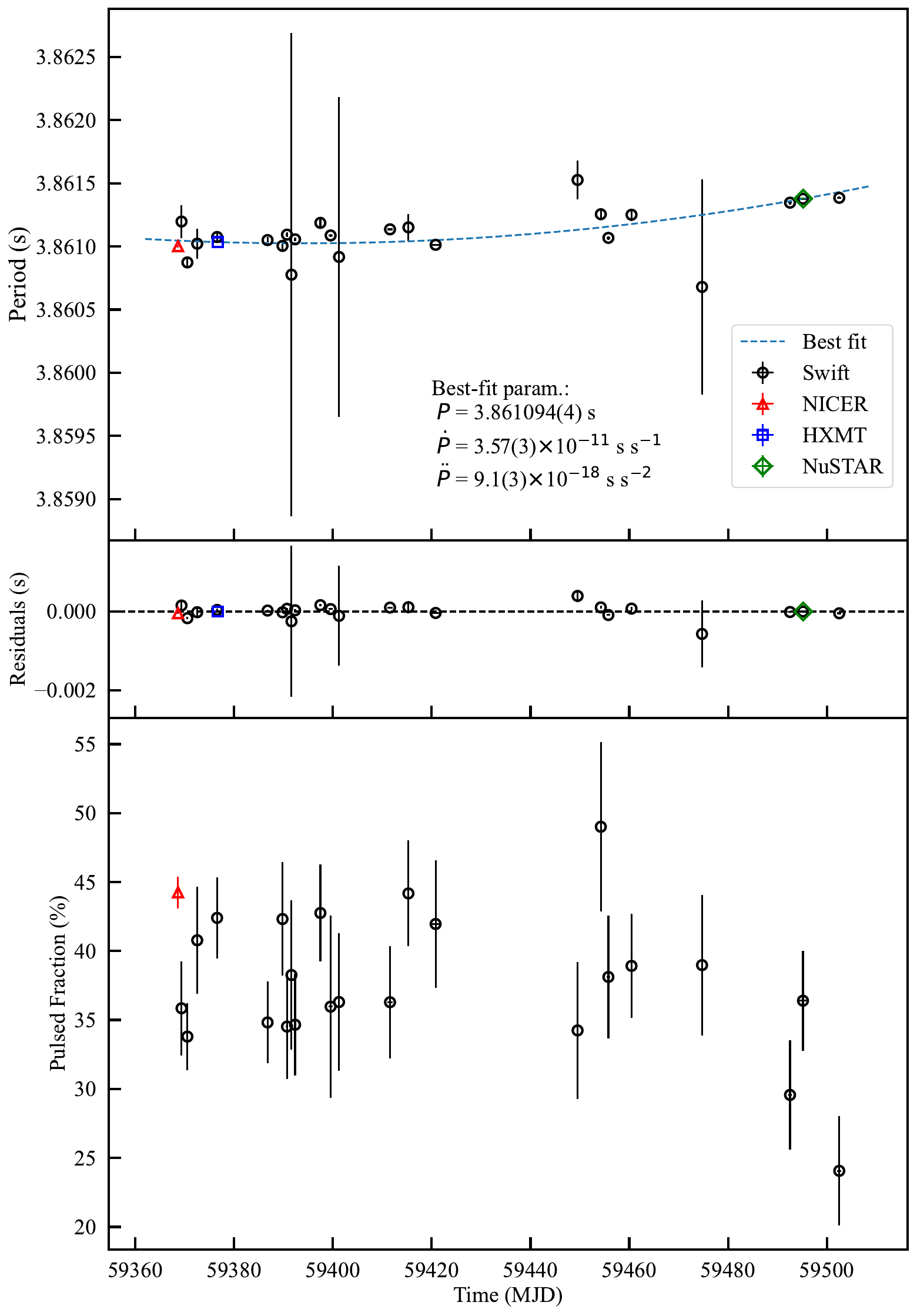}
\includegraphics[width=1.\columnwidth]{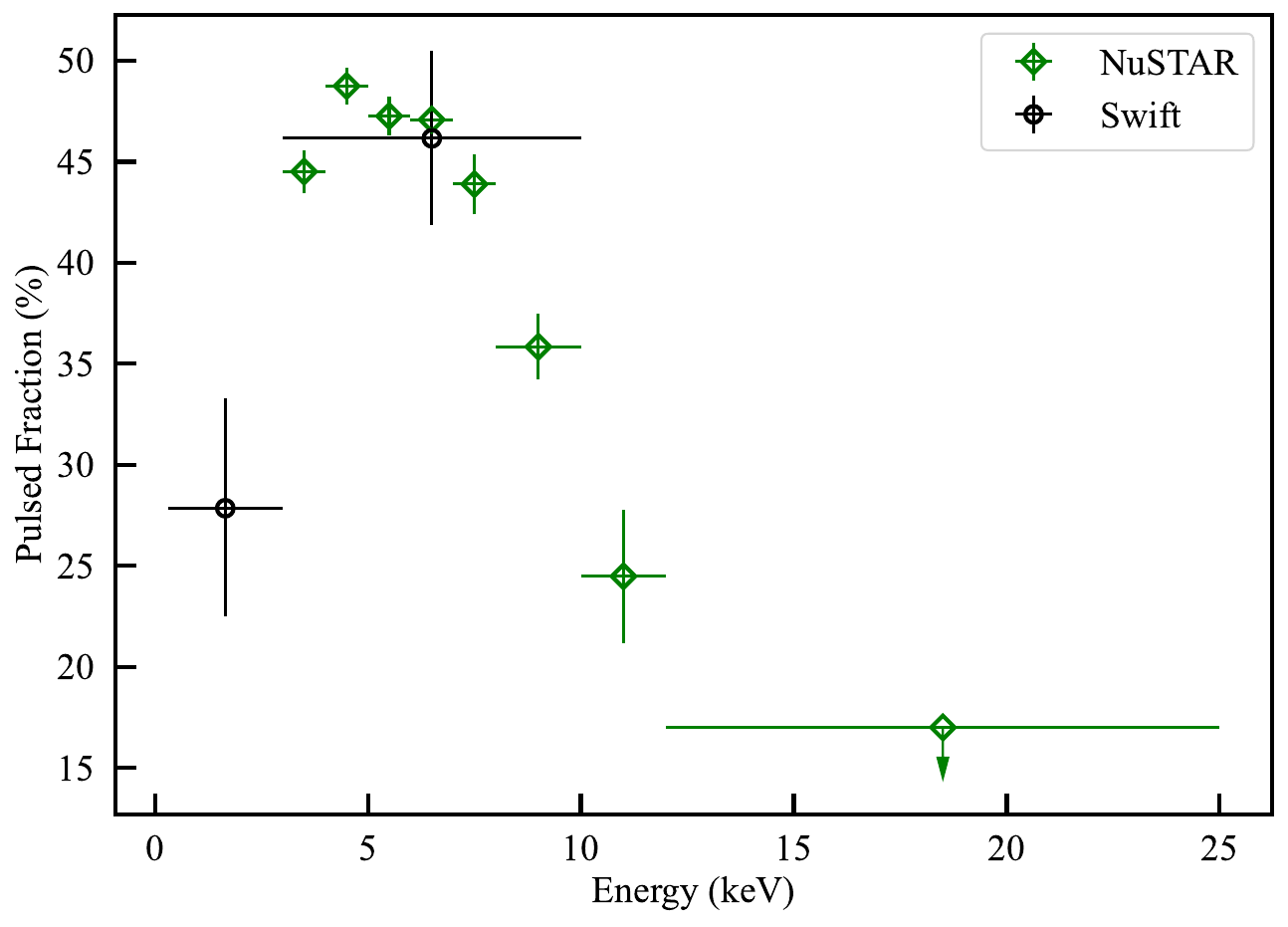}
\caption{{\it Top}: Values of the spin period measured in the single \nicer, \swift, \hxmt/LE and \nustar\ observations in 2021 as a function of time. The blue dashed line indicates the best-fitting model (for more details see Sect.\,\ref{sec:timing}). Post-fit residuals are shown at the bottom. {\it Middle}: Evolution of the background-subtracted pulsed fraction of \src\ as a function of time. {\it Bottom}: Background-subtracted pulsed fraction as a function of energy for the simultaneous \swift\ (black circle) and \nustar\ (green diamond) observations. The upper limit is reported at 3$\sigma$ c.l..}
\label{fig:timing_period}
\end{figure}

A phase-connected timing solution covering the first month of the outburst has already been reported by \cite{enoto21b}. Given the sparse cadence of our subsequent \swift\ observations up to 2021 October 22 (i.e. before the occurrence of a 2.5-month observation gap due to Solar constraints), we did not attempt to extrapolate that timing solution. Instead, we used it as a basis for determining an initial trial period at each epoch of our observations. We then employed a phase-fitting technique to calculate a more precise value of the period for each observation in which pulsations were detected with high significance. We selected events in the 0.3--10\,keV energy band for \swift, 1.5--8\,keV for \nicer, 2.5--6.5\,keV for \hxmt/LE and 3--10\,keV for \nustar. For the latter, we combined the FPMA and FPMB event files. The period evolution obtained between 2021 June 3 and October 22 using this procedure is shown in the top panel of Figure\,\ref{fig:timing_period}. To constrain the average spin-down rate over this time span, we fit the period evolution with a Taylor series truncated to the third order. At a reference epoch 59438.9464 MJD (2021 August 12), which is the mid-point of the time span considered, we found $P$ = 3.861094(4)\,s, $\dot{P}$ = 3.57(3) $\times$ 10$^{-11}$\,s\,s$^{-1}$ and $\ddot{P}$ = 9.1(3) $\times$ 10$^{-18}$\,s\,s$^{-2}$. The high reduced chi-square value obtained from this modelling, \rchisq = 22 for 23 dof, is most likely indicative of the presence of strong timing noise, as was also observed in the high-cadence \nicer\ campaign during the first month of the outburst \citep{enoto21b}. 

From the beginning of 2022 until the end of our campaign, the spin signal was only barely detected in the single \swift\ pointings. Therefore, we proceeded to analyse only the \nicer\ data collected in 2022 July--August in an attempt to extract a phase-connected timing solution. 
Through epoch-folding techniques \citep{leahy83}, we first estimated a spin period $P=3.858(2)\,\mathrm{s}$ in observation IDs 5202190101-5202190102 (which were merged for the timing analysis). Starting from this value and using a phase-fitting technique, we derived a phase-connected timing solution covering observations from Obs.ID 5202190101 (59781 MJD, 2022 July 21) to Obs.ID 5202190111 (59806 MJD, 2022 August 15). We did not include Obs.ID 5202190112 as the spin signal is barely detected in this data set. For Obs. IDs 5202190106, 5202190110 and 5202190111, we only considered events in the energy range 2--7\,keV, 2--6\,keV and 1.8--7\,keV, respectively, due to the high background level. We derived the following timing solution: $P = 3.862065(1)\,\mathrm{s}$, $\dot{P}=2.1(1)\times10^{-11}$\,s\,s$^{-1}$ at a reference epoch $T_0=59781.0\,\mathrm{MJD}$ (2022 July 21). 

In all the observations where pulsations had been found, the pulse profile appeared single-peaked and quasi-sinusoidal. The middle panel of Figure\,\ref{fig:timing_period} shows the variability of the pulsed fraction (here defined as the semi-amplitude of the best-fitting sinusoidal function to the profile divided by the source average net count rate) as a function of time as derived using the \nicer\ and \swift/XRT data sets in 2021. The pulsed fraction did not display significant variation over time, remaining consistently within the range $\sim$30–45\% during our observations until October 15 (last observation where pulsations were significantly detected during 2021), when it dropped to $\sim$24\%. No significant variability was also observed in the pulsed fraction of the \nicer\ pulse profiles acquired in 2022.
For the only epoch with a simultaneous \nustar\ observation, we detected pulsed emission up to $\sim$12\,keV and derived a 3$\sigma$ upper limit of $\sim$17\% for the pulsed fraction in the 12--25\,keV energy band. The pulsed fraction increased from $\sim$27\% in the softer band (0.3--3\,keV) to a maximum of $\sim$46\% around 5--6\,keV and then decreased with energy to $\sim$24\%, as shown in the bottom panel of Figure\,\ref{fig:timing_period}. The same trend was already observed by \cite{enoto21b} during the very early stage of the outburst. 

\begin{figure}
\includegraphics[width=1.7\columnwidth]{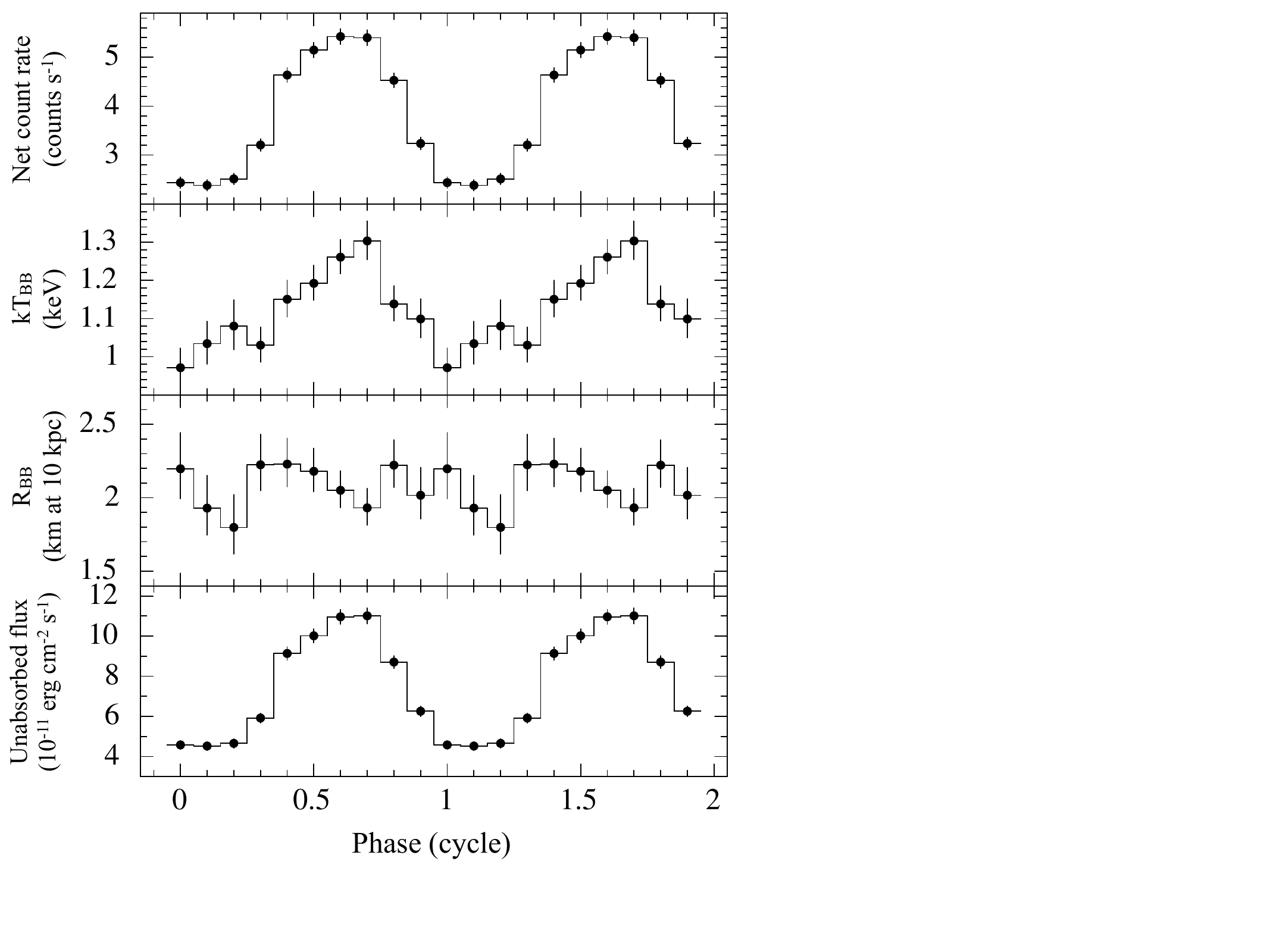} 
\vspace{-1.5cm}
\caption{\label{fig:pps} From top to bottom: background-subtracted pulse profile extracted from \nicer\ data at the outburst peak in the 1.5--8\,keV energy range, blackbody temperature and radius (assuming a distance of 10\,kpc), 0.3--10\,keV unabsorbed flux. 
All uncertainties are at 1$\sigma$ c.l. For display purposes, the pulse profile has been shifted arbitrarily in phase, and two cycles are shown.}
\end{figure}

We performed a phase-resolved spectral analysis using the \nicer\ data set acquired at the outburst onset. We divided the rotational phase cycle into 10 intervals, each of width 0.1 in phase, and fit an absorbed blackbody model to each phase-resolved spectrum. In the fits, \nh\ was held fixed to the phase-averaged value, while all other parameters were allowed to vary. The overall fit quality was good, with $\chi^2_r = 0.98$ for 177 dof. The results are shown in Figure\,\ref{fig:pps}. The modulation of the X-ray emission along the phase can be ascribed mainly to variations in the blackbody temperature, while the size of the emitting region seems to remain steady (fitting a constant term to the evolution of the blackbody radius as a function of phase gives $R_{{\rm BB}}$ = 2.07$\pm$0.05\,km and $\chi^2_r$ = 0.66 for 9 dof). Tying up the blackbody radius across the phases resulted in an equally satisfactory fit, with $\chi^2_r = 1.02$ for 186 dof. On the other hand, linking the blackbody temperature across the phases led to a slightly worse fit, with $\chi^2_r = 1.20$ for 186 dof. 
We also performed phase‐resolved spectral analysis on the 2022 \nicer\ dataset. Due to the significantly lower count rate, we divided the rotational cycle into three equal phase bins to extract statistically meaningful spectra. In each bin, we fitted an absorbed blackbody model, keeping $N_{\rm H}$ fixed to the phase‐averaged value and allowing the blackbody temperature and radius to vary. All blackbody parameters in the three phase intervals were consistent with one another at the 2$\sigma$ confidence level. This indicates no statistically significant phase‐dependent variation in either temperature or emitting area in the 2022 data.

\subsection{Search for short X-ray bursts}
\label{sec:bursts}

We used the \integral\ archival data to look for short bursts from \src. However, during most pointings, \src\  was at large off-axis angles in the partially coded field of view of IBIS, which reduced the sensitivity for burst detection. The burst search was performed by selecting events from the pixels that were illuminated by \src\ for at least 50\% of their area. We screened the light curves in the nominal 15-150\,keV energy range binned at eight logarithmically spaced time resolutions between 0.01 to 1.28\,s.
Potential triggers were selected based on threshold levels corresponding to a chance occurrence of $10^{-3}$ in each ScW. 
After grouping together triggers at different timescales that belonged to the same event, we carried out an imaging analysis of each candidate.
None of them could be confirmed as a point source event coming from the direction of \src.
Therefore, we can conclude that no bursts from \src\ were detected, with a typical upper limit of $\sim2\times10^{-8}$ erg\,cm$^{-2}$ on the fluence in the 30--150\,keV energy range (the exact value depends on the off-axis angle and background level in each ScW).

We also conducted a search for short bursts on \nicer, \swift\ and \nustar\ light curves adopting different timing resolutions (2$^{-4}$, 2$^{-5}$, 2$^{-6}$, 2$^{-7}$ and 2$^{-8}$\,s), except for the \swift/XRT PC-mode event files that were binned at the available timing resolution (2.5073\,s). We tagged as part of a burst every bin with a probability lower than 10$^{-4}$($N N_{\rm trials}$)$^{-1}$ of being a random fluctuation compared to the average count rate of the full observation, considering the total number of bins $N$ and the timing resolutions $N_{\rm trials}$ used in the search. We identified a total of 56 bursts, whose times of arrival are listed in Table\,\ref{tab:bursts}. Due to the low photon counting statistics, it was not possible to perform a meaningful spectral analysis of such events. During our monitoring campaign, \src\ emitted a few short bursts that triggered the X-ray all-sky monitors (e.g., \swift/BAT, \citealt{enoto21b}; GECAM, \citealt{zhang2022_gcn}). We did not find any bursts in the \swift/XRT and \nicer/XTI data sets simultaneous to the reported events.

\subsection{Radio searches}
\label{sec:radio}

The \pks\ data were folded using the best X-ray ephemeris (Sect.\,\ref{sec:timing}) and searched over a period range spanning $\pm 300$\,\textmu s around the spin period at the epoch of the radio observations, and over a dispersion measure range from 0 to 1500\,pc\,cm$^{-3}$. UWL data were also split into three sub-bands, 0.7--1\,GHz, 1--2\,GHz and 2--4\,GHz, and folded independently. No periodic pulsations at the X-ray period were found in either data set. The derived flux density upper limits for the three sub-bands are reported in Table\,\ref{tab:obsradio}. We do not report values for the PDFB4 data, since they were acquired over smaller bandwidths and thus all the resulting upper limits are significantly worse than those from the simultaneous data taken with the {\it Medusa} backend. 

The single-pulse search for both backends was carried out via the SPANDAK\footnote{\url{https://github.com/gajjarv/PulsarSearch}} \citep{gajjar18} pipeline. The initial RFI purging is performed by SPANDAK through {\sc rfifind} from the {\sc PRESTO}\footnote{\url{https://www.cv.nrao.edu/~sransom/presto/}} package. The pipeline automates the search for bursts/single pulses using {\sc Heimdall} \citep{barsdell12}, which performs a single pulse search by exploiting the matched filtering technique along a range of trial DMs. The search in DM of candidates run in the range from 0 to 2000\,pc\,cm$^{-3}$ and only candidates generated by {\sc Heimdall} with S/N $\geq$ 8 and pulse width in the range 0.128--65.536\,ms have been selected. The search was carried out over both the full bandwidth (3328\,MHz) by dividing that into sub-bands (1664\,MHz, 832\,MHz, and 416\,MHz wide) and processing them individually (see e.g. \citealt{kumar21}). This was done to increase the sensitivity for single pulses with a spectral width smaller than the whole UWL bandwidth. We also considered additional sub-bands, shifted in frequency by half sub-band for each of the mentioned explored cases. For data taken with the PDFB4 backend, it was not necessary to perform a sub-band search for candidates, given the smaller bandwidth. All candidates were also visually scrutinised to cross-check potential candidates found in each backend. No robust candidate possessing the characteristics of a fast radio burst was found with upper limits listed in Table\,\ref{tab:obsradio}.

All the X-ray bursts on June 7 fell within the radio observation windows, and the same held true for all but the first burst on June 4 and for all but the last three bursts on June 5. Hence, the dedispersed radio time-series were closely inspected (after a detailed manual cleaning) in a range of 30\,s around the times of the occurrence of the X-ray bursts. No feature resembling a burst was observed down to S/N=6, implying improved upper limits with respect to single pulse searches performed over the whole dataset (see Table\,\ref{tab:obsradio}).

\section{Discussion}
\label{sec:discussion}

We presented a comprehensive study of the first recorded outburst of the magnetar \srcfirst, based on an extensive campaign of X-ray observations spanning nearly 29 months, and accompanied by deep radio searches. This source displays a suite of remarkable observational characteristics: a very slow flux decay in the first phases, followed by a fast dimming, a persistently high blackbody temperature with no variation over hundreds of days, and a spin-down evolution characterised by substantial variability. In the following, we place these findings in the broader context of magnetar phenomenology and theoretical modelling. \\

{\bf A prolonged bright plateau.}
The persistent X-ray emission from \src\ showed a remarkable evolution throughout our campaign. Over the first three weeks,  the observed flux remained almost constant at $\sim$3.5$\times$10$^{-11}$\,\flux\ and only in the final months, it exhibited a more rapid drop, declining to $\sim$9$\times$10$^{-13}$\,\flux\ by 2023 October. The overall light curve is well described by an exponential with $e$-folding timescale of $\sim$260 days.
This prolonged plateau phase is unusual but not unprecedented among magnetars. For example, \sgrg\ and \ee\ also displayed delayed flux decays following their 2010 and 2016 outbursts, respectively \citep[][and reference therein]{cotizelati18}. However, the case of \src\ appears even more extreme, as the flux remained above $10^{-11}$\,\flux\ for more than 500 days -- among the longest such intervals ever observed.

This evolution raises the question of whether the true outburst onset may have preceded the first detected burst on 2021 June 3. Although we cannot exclude this possibility, the high flux level observed after the activation -- corresponding to a luminosity of $(2$–$8)\times10^{35}$\,\lum\ for a distance of 5--10\,kpc -- is comparable to the typical outburst peaks seen in magnetars. Moreover, this luminosity is close to the theoretical saturation limit imposed by neutrino cooling in the crust (as explored in early works \citealt{pons12} and recently reassessed in the cooling simulations by \citealt{2025A&A...701A.229D}) as well as to the maximum value predicted by twisted magnetosphere models \citep{beloborodov09}. These findings suggest that the observed plateau likely reflects the actual early evolution of the outburst.\\

{\bf A constant temperature and an evolving emitting area.}
A striking feature of the outburst from \src\ is the constancy of its thermal spectral properties over a prolonged interval. The blackbody temperature remained approximately constant at $kT_{\rm BB} \sim 1.2$\,keV for over two years, while the emitting radius declined from $\sim$1.7\,km to $\sim$0.3\,km. This behaviour is consistent with the gradual contraction of a hot spot on the neutron star surface as the outburst decay and is similar to that seen in the outbursts of other high-temperature magnetars. In particular, both \sgrl\ and \coronamag\ exhibited similarly hot thermal components, with $kT\gtrsim1$\,keV, which remained stable for extended periods, during their latest outburst. In the case of \sgrl, \citet{younes22} reported a double blackbody spectrum (with components at $\simeq$0.5 and 1.2\,keV) that persisted for more than 220 days without significant evolution. For \coronamag, \citet{hu20} found that the thermal flux decayed by $\simeq$60\% over $\simeq$100\,days, with only a modest decrease in emitting radius, and the temperature remaining at around 1.1\,keV as of late 2021 \citep{ibrahim24}.
Another notable example is the 2017 outburst of \wessy, during which the inferred blackbody temperature attained a high constant value of $\sim$0.7\,keV over $\sim$350 days \citep{borghese19}. Therefore, \src\ appears to belong to a growing subset of magnetars that are able to maintain a hot, small emitting region over long periods during outburst.
This phenomenology points to sustained heating mechanisms, potentially linked to magnetospheric currents continuously depositing energy on localised regions of the neutron star surface. The nearly constant temperature, coupled with a shrinking area, is consistent with the gradual contraction of a heated spot rather than bulk cooling of the entire crust (see below).

Near the outburst peak, the flux modulation at the spin period is primarily driven by temperature variations, with little to no variation in the emitting area. This suggests that the hot region has a complex, non-uniform temperature distribution. Such temperature gradients could arise from localised heating by returning currents along twisted magnetic field bundles, possibly elongated or fan-shaped, and are expected in twisted magnetosphere models \citep{beloborodov09}. An asymmetric thermal configuration can also appear in response to impulsive energy deposition within the star crust as the result of highly anisotropic heat transport to the surface in the presence of a complex magnetic field, as shown using 3D simulations \citep{degrandis22}. Similar behaviour has been observed in, e.g., \sgrl\ \citep{cotizelati21_sgr1830, younes22} and \xte\ \citep{borghese21}.
In contrast, the 2022 \nicer\ observations revealed no statistically significant phase-dependent variation in either temperature or radius. This likely reflects the lower photon statistics as well as a more compact and stable emitting region at later stages of the outburst. \\

{\bf Timing evolution and spin-down behaviour.}
Our timing analysis provides a view of the rotational evolution of \src\ over the first five months and a snapshot from the 2022 \nicer\ observations. The period derivative $\dot{P} \sim 3.6 \times 10^{-11}$\,s\,s$^{-1}$ measured during the first five months is slightly ($\sim$15\%) higher than that obtained during the first month by \citet{enoto21b}, suggesting a relatively stable spin-down torque during this phase. However, the data show evidence of timing noise, as testified by the presence of residuals in the fits with a polynomial.

Using the standard vacuum dipole formula, we estimate a surface dipolar magnetic field strength of $B_{\rm dip}\simeq7.5 \times 10^{14}$\,G, a characteristic age $\tau_{\rm c}\simeq1.7$\,kyr, and a spin-down luminosity $\dot{E}_{\rm rot}\simeq2.5 \times 10^{34}$\,\lum. These parameters place \src\ among the strongly magnetised and relatively young members of the magnetar population.

The \nicer\ timing campaign in mid-2022, almost a year after the outburst onset, yielded a lower spin-down rate, $\dot{P}\simeq 2.1 \times 10^{-11}$\,s\,s$^{-1}$. This likely indicates that the magnetosphere was relaxing back toward its pre-outburst configuration. Large changes in torque are commonly observed in magnetars during outbursts (e.g., \citealt{dib12,scholz17,archibald20,rajwade22}) and are generally attributed to the evolution of currents in twisted magnetospheres (see below for a discussion of such a scenario).\\

{\bf X-ray bursts and radio silence.}
We identified 56 short X-ray bursts in our observations of \src, primarily during the early months of the outburst. This bursting activity is consistent with the behaviour of many magnetars during active phases. 

No radio pulsations or bursts were detected in our three \pks/Murriyang observations, either as periodic signals or as bursts coincident with X-ray bursts. This confirms the findings of \citet{enoto21b}, who also reported no radio emission from \src\ using multiple telescopes. While a few active magnetars, including \sgrf\ and \coronamag, have shown transient radio pulsations during or shortly after outbursts, not all magnetars are radio-loud. The absence of detectable radio emission from \src\ may result from unfavourable viewing geometry, intrinsic emission properties, or absorption in the magnetosphere or local environment.\\

\begin{figure}
    \centering
    \includegraphics[width=0.95\linewidth]{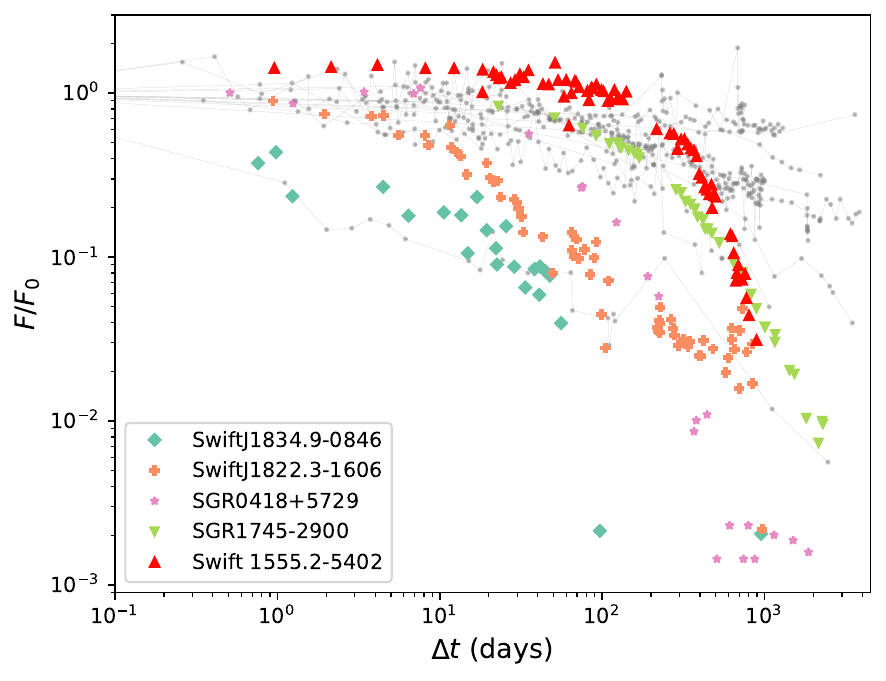}
    \caption{\label{fig:mooc} Comparison between the long-term temporal evolution of the normalised flux for the major magnetar outbursts occurred up to the end of 2016 with the outburst of \src\ in red. Data taken from the Magnetar Outburst Online Catalog \citep{cotizelati18}. For \galmag, we included the latest observations presented by \citealt{rea20}. The outbursts highlighted in colour are those showing a substantial and fast decay of the flux in the final phase, as observed for \src. }\label{fig:comparison}
\end{figure}

{\bf Implications for outburst mechanisms.}
The outburst of \src\ challenges the accepted paradigms in several respects.
In the initial phases, the emission kept a high luminosity for an unusually long time span, with a nearly constant blackbody temperature. Subsequently, the temperature maintained a nearly constant value, with the flux and equivalent emission radii decaying following exponential profiles. The corresponding decay times are compatible with being one double the other, corresponding to a simple scenario in which the flux varies only due to the shrinking of the emission area.
Figure\,\ref{fig:comparison} shows a comparison between the event at hand and the known outburst sample as described by \citet{cotizelati18}. Since the distance to this source is unknown, we normalised all the curves to the flux level at the outburst onset. Moreover, we highlight four other cases that display a similar morphology, namely, a rather stable initial phase followed by a substantial, rapid flux decay. Not only \src\ exhibits a very extended initial plateau, but the subsequent drop phase is much steeper than in other sources.

\begin{figure}
    \centering \includegraphics[width=0.95\linewidth]{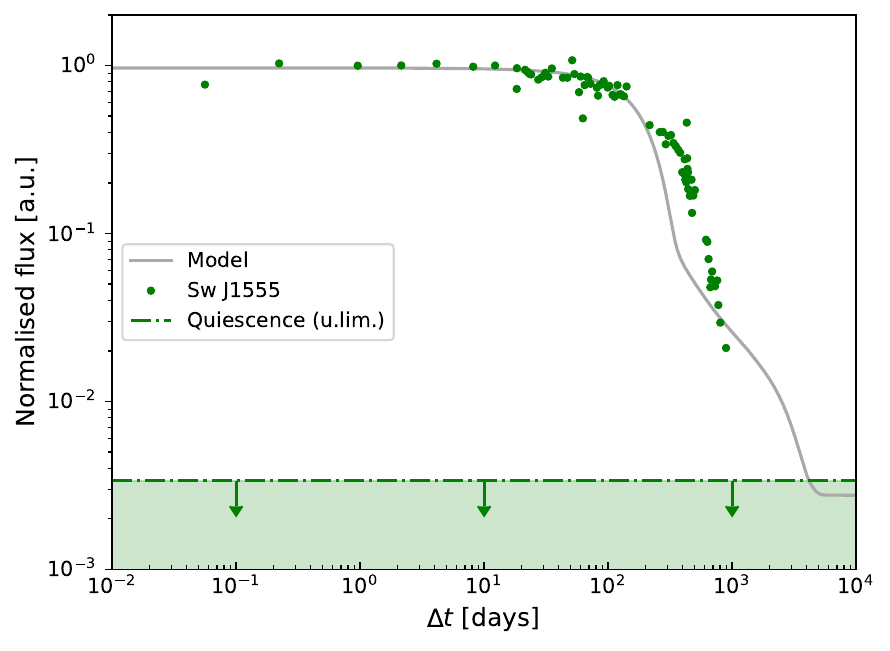}
    \caption{Comparison between the flux evolution of \src\ and a model of hotspot cooling modelled with the methodology by \citet{2025A&A...701A.229D}, showing the evolution of a $\sim10^{44}\,$erg injection in a 3\,km radius hotspot over a time of $\sim400\,$days.}
    \label{fig:model}
\end{figure}

Figure\ \ref{fig:model} shows a comparison between the data and a model of cooling of a crustal hotspot obtained with the methodology presented by \citet{2025A&A...701A.229D}, which describes the evolution of the luminosity after an arbitrary amount of heat has been located in a small portion of the crust for a certain time. In order to reproduce the behaviour of the data (although performing a formal fit over the several model parameters is beyond the scope of this work), and in particular the long plateau phase, the time during which the heat is injected must be rather long, of the same order of the plateau itself (specifically, 400 days in the shown example). This implies either that the heat dissipation mechanism is long-lived, or that a sequence of multiple shorter heating episodes is active for an extended duration (see also \citealt{pons12}). Still, we must note that, even though the model successfully describes the plateau phase, it cannot reproduce the features of the decay apart from the overall timescale, suggesting that several physical ingredients are still missing.
Moreover, neutrino emission in the crust \citep{yakovlev01} limits the maximum temperature that can be reached in the outermost layers (the so-called envelope). This was realised by \citet{potekhinponspage} and more recently reassessed by \citep{2025A&A...701A.229D}, \citep{kovlakas2025}, who showed that considering a thin envelope (the appropriate assumption when treating short-term phenomena), this maximum temperature is higher than that sustainable in a thick envelope, reaching values of about $1-2\times10^7$\,K. This would match the observed value of 1.2\,keV, corresponding to a surface temperature $\sim 1.7\times10^7\,$K accounting for the gravitational redshift on the surface of a standard neutron star with $M_{\rm{NS}}=1.4\,\text{M}_\odot$ and $R_{\rm{NS}}=12\,$km. 

An alternative (or complementary) explanation might involve prolonged magnetospheric currents in the spirit of twisted bundle models \citep{beloborodov09}, whereby a slowly dissipating bundle of currents maintains surface heating over an extended period. If this twist is sustained or replenished, it could naturally account for the observed prolonged brightness along with the temperature gradients inferred from phase-resolved spectra, which point to a complex and possibly evolving magnetospheric configuration. In this framework, the initially twisted magnetic bundle inflates field lines and enhances the open‐field region, thereby increasing the spin‐down torque. As the twist dissipates, the radius of the hot region should shrink, and the torque should decrease monotonically. However, the variability seen in \src, and in other cases like \ee\ and \coronamag, suggests a more complex scenario -- possibly involving the formation and dissipation of multiple, rapidly evolving twisted regions. These could generate spatially and temporally variable currents, leading to torque fluctuations that are not strictly monotonic. 
Future modelling efforts should therefore explore whether scenarios involving multiple, evolving twists can simultaneously reproduce both the flux and torque evolution observed in \src.

Furthermore, outbursts have been modeled within the fallback disk paradigm \citep[e.g.,][]{2003ApJ...593L..93E} by \citet{2012ApJ...758...98C}. They found that the luminosity of such events decays over the first $\sim 100$ days as $\propto t^{-n}$, with $0.5 \lesssim n \lesssim 1$. This behavior contrasts with both the remarkably flat evolution observed during the early phases of the event considered here and the subsequent steep decline (that would correspond to $n \gtrsim 2$). Therefore, the observed flux evolution of the outburst \src\ is difficult to reconcile with this scenario.

\section{Conclusions}
We studied the long-term evolution of the X-ray properties of the magnetar \srcfirst\ during the first $\sim$29 months of its first (recorded) outburst, and analysed three radio observations carried out soon after the event onset without any successful detection. 

The peculiarity of this outburst is its temporal evolution characterised by a long bright plateau of about 500 days followed by a sharp, fast decline (see Fig.\,\ref{fig:comparison}). Deep observations at late time are required to understand whether the source flux continued to decay below the quiescence upper limit or it has already reached its (unknown) quiescent level. An additional puzzling trait is the blackbody temperature of $\sim$1.2\,keV that remained constant during the entire duration of the monitoring campaign ($\sim$29 months). These features represent a challenge for the outburst mechanisms proposed so far. While crustal cooling models are able to reproduce the plateau phase, they fail to model the fast decay in the final phase. On the other hand, detailed models concerning the un-twisting of magnetic field bundles in the magnetosphere are still missing.

\begin{acknowledgements}
We thank K.\ Gendreau for approving our \nicer\ Target of Opportunity (ToO) requests and B.\ Cenko and the \swift\ duty scientists and science planners for making the \swift\ ToO observations possible. We thank the \hxmt\ Science Operations Centre for promptly scheduling our ToO program. 
We thank the director of the \pks/Murriyang Observatory for the prompt scheduling of our NAPA proposal.
This research is based on observations with \swift\ (NASA/ASI/UKSA), \nicer\ (NASA), \nustar\ (CalTech/NASA/JPL), \hxmt\ (CNSA, CAS), \integral\ (ESA) and on data retrieved through the NASA/GSFC's HEASARC archives. 
We used data collected at the \pks/Murriyang radio telescope (proposal No.\ P1083), part of the Australia Telescope National Facility which is funded by the Australian Government for operation as a National Facility managed by CSIRO. We acknowledge the Wiradjuri people as the traditional owners of the Observatory site.\\
A.B.\ acknowledges support through the European Space Agency (ESA) research fellowship programme.
F.C.Z.\ is supported by a Ram\'on y Cajal fellowship (grant agreement RYC2021-030888-I). D.D.G.\ is supported by a Juan de la Cierva fellowship (grant agreement JDC2023-052264-I). F.C.Z.\ M.I.\ D.D.G.\ and N.R.\ are supported by the ERC Consolidator Grant ``MAGNESIA" (No.\ 817661) and acknowledge funding from the Catalan grant SGR2021-01269 (PI: Graber/Rea), the Spanish grant ID2023-153099NA-I00 (PI: Coti Zelati), and by the program Unidad de Excelencia Maria de Maeztu CEX2020-001058-M. M.I.\ is also supported by the ERC Proof of Concept ``DeepSpacePulse" (No. 101189496). G.L.I. and P.E. acknowledge financial support from the Italian MIUR through PRIN grant 2017LJ39LM. G.L.I.\ also acknowledges funding from ASI-INAF agreements I/037/12/0 and 2017-14-H.O.

S.M.\ and D.P.P.\ acknowledge financial support through the INAF grants ``Magnetars'' and ``Toward Neutron Stars Unification''.
L.D. acknowledges funding from German Research Foundation (DFG), Projektnummer 549824807.
M.B, S.M., M.P., and A.P.\ acknowledge resources from the research grant ``iPeska'' (PI: Possenti), funded under the INAF national call Prin-SKA/CTA approved with the Presidential Decree 70/2016.

This work was also partially supported by the program Unidad de Excelencia Mar\'ia de Maeztu CEX2020-001058-M. 
This work has been partially supported by the ASI-INAF program I/004/11/5 and 
\end{acknowledgements}

\bibliographystyle{aa}
\bibliography{biblio}

\begin{thebibliography}{57}
\expandafter\ifx\csname natexlab\endcsname\relax\def\natexlab#1{#1}\fi

\bibitem[{{Archibald} {et~al.}(2020){Archibald}, {Scholz}, {Kaspi},
  {Tendulkar}, \& {Beardmore}}]{archibald20}
{Archibald}, R.~F., {Scholz}, P., {Kaspi}, V.~M., {Tendulkar}, S.~P., \&
  {Beardmore}, A.~P. 2020, \apj, 889, 160

\bibitem[{{Arnaud}(1996)}]{arnaud96}
{Arnaud}, K.~A. 1996, in Astronomical Data Analysis Software and Systems V,
  Vol. 101, XSPEC: The First Ten Years, ed. G.~H. {Jacoby} \& J.~{Barnes} (ASP,
  San Francisco), 17--20

\bibitem[{{Barsdell} {et~al.}(2012){Barsdell}, {Bailes}, {Barnes}, \&
  {Fluke}}]{barsdell12}
{Barsdell}, B.~R., {Bailes}, M., {Barnes}, D.~G., \& {Fluke}, C.~J. 2012,
  MNRAS, 422, 379

\bibitem[{{Beloborodov}(2009)}]{beloborodov09}
{Beloborodov}, A.~M. 2009, \apj, 703, 1044

\bibitem[{{Bernardini} {et~al.}(2021){Bernardini}, {Gropp}, {Kennea},
  {Klingler}, {Kuin}, {Laha}, {Lien}, {Melandri}, {Page}, {Palmer}, {Sbarrato},
  {Sbarufatti}, {Siegel}, {Tohuvavohu}, \& {Neil Gehrels Swift Observatory
  Team}}]{bernardini21}
{Bernardini}, M.~G., {Gropp}, J.~D., {Kennea}, J.~A., {et~al.} 2021, GRB
  Coordinates Network, 30296, 1

\bibitem[{{Borghese} {et~al.}(2019){Borghese}, {Rea}, {Turolla}, {Pons},
  {Esposito}, {Coti Zelati}, {Savchenko}, {Bozzo}, {Perna}, {Zane},
  {Mereghetti}, {Campana}, {Mignani}, {Bachetti}, {Rodr{\'\i}guez}, {Pintore},
  {Tiengo}, {G{\"o}tz}, {Israel}, \& {Stella}}]{borghese19}
{Borghese}, A., {Rea}, N., {Turolla}, R., {et~al.} 2019, \mnras, 484, 2931

\bibitem[{{Borghese} {et~al.}(2021){Borghese}, {Rea}, {Turolla}, {Rigoselli},
  {Alford}, {Gotthelf}, {Burgay}, {Possenti}, {Zane}, {Coti Zelati}, {Perna},
  {Esposito}, {Mereghetti}, {Vigan{\`o}}, {Tiengo}, {G{\"o}tz}, {Ibrahim},
  {Israel}, {Pons}, \& {Sathyaprakash}}]{borghese21}
{Borghese}, A., {Rea}, N., {Turolla}, R., {et~al.} 2021, \mnras, 504, 5244

\bibitem[{{Burrows} {et~al.}(2005){Burrows}, {Hill}, {Nousek}, {Kennea},
  {Wells}, {Osborne}, {Abbey}, {Beardmore}, {Mukerjee}, {Short}, {Chincarini},
  {Campana}, {Citterio}, {Moretti}, {Pagani}, {Tagliaferri}, {Giommi},
  {Capalbi}, {Tamburelli}, {Angelini}, {Cusumano}, {Br{\"a}uninger}, {Burkert},
  \& {Hartner}}]{burrows05}
{Burrows}, D.~N., {Hill}, J.~E., {Nousek}, J.~A., {et~al.} 2005, Space Science
  Reviews, 120, 165

\bibitem[{{Carrasco} {et~al.}(2019){Carrasco}, {Vigan{\`o}}, {Palenzuela}, \&
  {Pons}}]{carrasco2019}
{Carrasco}, F., {Vigan{\`o}}, D., {Palenzuela}, C., \& {Pons}, J.~A. 2019,
  \mnras, 484, L124

\bibitem[{{{\c{C}}al{\i}{\c{s}}kan} \& {Ertan}(2012)}]{2012ApJ...758...98C}
{{\c{C}}al{\i}{\c{s}}kan}, {\c{S}}. \& {Ertan}, {\"U}. 2012, \apj, 758, 98

\bibitem[{{Collazzi} {et~al.}(2015){Collazzi}, {Kouveliotou}, {van der Horst},
  {Younes}, {Kaneko}, {G{\"o}{\u{g}}{\"u}{\textcommabelow s}}, {Lin}, {Granot},
  {Finger}, {Chaplin}, {Huppenkothen}, {Watts}, {von Kienlin}, {Baring},
  {Gruber}, {Bhat}, {Gibby}, {Gehrels}, {McEnery}, {van der Klis}, \&
  {Wijers}}]{collazzi15}
{Collazzi}, A.~C., {Kouveliotou}, C., {van der Horst}, A.~J., {et~al.} 2015,
  \apjs, 218, 11

\bibitem[{{Cordes} \& {Lazio}(2002)}]{ne2001}
{Cordes}, J.~M. \& {Lazio}, T.~J.~W. 2002, arXiv e-prints, astro

\bibitem[{{Coti Zelati} {et~al.}(2021{\natexlab{a}}){Coti Zelati}, {Borghese},
  {Israel}, {Rea}, {Esposito}, {Pilia}, {Burgay}, {Possenti}, {Corongiu},
  {Ridolfi}, {Dehman}, {Vigan{\`o}}, {Turolla}, {Zane}, {Tiengo}, \&
  {Keane}}]{cotizelati21_sgr1830}
{Coti Zelati}, F., {Borghese}, A., {Israel}, G.~L., {et~al.}
  2021{\natexlab{a}}, \apjl, 907, L34

\bibitem[{{Coti Zelati} {et~al.}(2021{\natexlab{b}}){Coti Zelati}, {Borghese},
  {Rea}, {Israel}, {Esposito}, {Enoto}, \& {Campana}}]{cotizelati21}
{Coti Zelati}, F., {Borghese}, A., {Rea}, N., {et~al.} 2021{\natexlab{b}}, The
  Astronomer's Telegram, 14674, 1

\bibitem[{{Coti Zelati} {et~al.}(2018){Coti Zelati}, {Rea}, {Pons}, {Campana},
  \& {Esposito}}]{cotizelati18}
{Coti Zelati}, F., {Rea}, N., {Pons}, J.~A., {Campana}, S., \& {Esposito}, P.
  2018, \mnras, 474, 961

\bibitem[{{De Grandis} {et~al.}(2025){De Grandis}, {Rea}, {Kovlakas}, {Coti
  Zelati}, {Vigan{\`o}}, {Ascenzi}, {Pons}, {Turolla}, \&
  {Zane}}]{2025A&A...701A.229D}
{De Grandis}, D., {Rea}, N., {Kovlakas}, K., {et~al.} 2025, \aap, 701, A229

\bibitem[{{De Grandis} {et~al.}(2022){De Grandis}, {Turolla}, {Taverna},
  {Lucchetta}, {Wood}, \& {Zane}}]{degrandis22}
{De Grandis}, D., {Turolla}, R., {Taverna}, R., {et~al.} 2022, \apj, 936, 99

\bibitem[{{Dehman} {et~al.}(2020){Dehman}, {Vigan{\`o}}, {Rea}, {Pons},
  {Perna}, \& {Garcia-Garcia}}]{dehman2020}
{Dehman}, C., {Vigan{\`o}}, D., {Rea}, N., {et~al.} 2020, \apjl, 902, L32

\bibitem[{{Dib} {et~al.}(2012){Dib}, {Kaspi}, {Scholz}, \& {Gavriil}}]{dib12}
{Dib}, R., {Kaspi}, V.~M., {Scholz}, P., \& {Gavriil}, F.~P. 2012, \apj, 748, 3

\bibitem[{{Enoto} {et~al.}(2021){Enoto}, {Ng}, {Hu}, {G{\"u}ver}, {Jaisawal},
  {O'Connor}, {G{\"o}{\u{g}}{\"u}{\c{s}}}, {Lien}, {Kisaka}, {Wadiasingh},
  {Majid}, {Pearlman}, {Arzoumanian}, {Bansal}, {Blumer}, {Chakrabarty},
  {Gendreau}, {Ho}, {Kouveliotou}, {Ray}, {Strohmayer}, {Younes}, {Palmer},
  {Sakamoto}, {Akahori}, \& {Eie}}]{enoto21b}
{Enoto}, T., {Ng}, M., {Hu}, C.-P., {et~al.} 2021, \apjl, 920, L4

\bibitem[{{Ertan} \& {Alpar}(2003)}]{2003ApJ...593L..93E}
{Ertan}, {\"U}. \& {Alpar}, M.~A. 2003, \apjl, 593, L93

\bibitem[{Esposito {et~al.}(2021)Esposito, Rea, \& Israel}]{esposito21}
Esposito, P., Rea, N., \& Israel, G.~L. 2021, Magnetars: A Short Review and
  Some Sparse Considerations, ed. T.~M. Belloni, M.~M{\'e}ndez, \& C.~Zhang
  (Berlin, Heidelberg: Springer Berlin Heidelberg), 97--142

\bibitem[{{Evans}(2021)}]{evans21}
{Evans}, P.~A. 2021, The Astronomer's Telegram, 14675, 1

\bibitem[{{Gajjar} {et~al.}(2018){Gajjar}, {Siemion}, {Price}, {Law},
  {Michilli}, {Hessels}, {Chatterjee}, {Archibald}, {Bower}, {Brinkman},
  {Burke-Spolaor}, {Cordes}, {Croft}, {Enriquez}, {Foster}, {Gizani},
  {Hellbourg}, {Isaacson}, {Kaspi}, {Lazio}, {Lebofsky}, {Lynch}, {MacMahon},
  {McLaughlin}, {Ransom}, {Scholz}, {Seymour}, {Spitler}, {Tendulkar},
  {Werthimer}, \& {Zhang}}]{gajjar18}
{Gajjar}, V., {Siemion}, A.~P.~V., {Price}, D.~C., {et~al.} 2018, \apj, 863, 2

\bibitem[{{Gendreau} {et~al.}(2012){Gendreau}, {Arzoumanian}, \&
  {Okajima}}]{gendreau12}
{Gendreau}, K.~C., {Arzoumanian}, Z., \& {Okajima}, T. 2012, in Society of
  Photo-Optical Instrumentation Engineers (SPIE) Conference Series, Vol. 8443,
  Space Telescopes and Instrumentation 2012: Ultraviolet to Gamma Ray, ed.
  T.~{Takahashi}, S.~S. {Murray}, \& J.-W.~A. {den Herder}, 844313

\bibitem[{{Goldwurm} {et~al.}(2003){Goldwurm}, {David}, {Foschini}, {Gros},
  {Laurent}, {Sauvageon}, {Bird}, {Lerusse}, \& {Produit}}]{goldwurm03}
{Goldwurm}, A., {David}, P., {Foschini}, L., {et~al.} 2003, \aap, 411, L223

\bibitem[{{Grefenstette} {et~al.}(2025){Grefenstette}, {Bhargava}, {Fuerst},
  {Bachetti}, {Li}, \& {rmludlam}}]{brian_grefenstette_2025_14969199}
{Grefenstette}, B., {Bhargava}, Y., {Fuerst}, F., {et~al.} 2025,
  {NuSTAR/nustar-gen-utils: The one with Zenodo}

\bibitem[{{Harrison} {et~al.}(2013){Harrison}, {Craig}, {Christensen},
  {Hailey}, {Zhang}, {Boggs}, {Stern}, {Cook}, {Forster}, {Giommi},
  {Grefenstette}, {Kim}, {Kitaguchi}, {Koglin}, {Madsen}, {Mao}, {Miyasaka},
  {Mori}, {Perri}, {Pivovaroff}, {Puccetti}, {Rana}, {Westergaard}, {Willis},
  {Zoglauer}, {An}, {Bachetti}, {Barri{\`e}re}, {Bellm}, {Bhalerao},
  {Brejnholt}, {Fuerst}, {Liebe}, {Markwardt}, {Nynka}, {Vogel}, {Walton},
  {Wik}, {Alexander}, {Cominsky}, {Hornschemeier}, {Hornstrup}, {Kaspi},
  {Madejski}, {Matt}, {Molendi}, {Smith}, {Tomsick}, {Ajello}, {Ballantyne},
  {Balokovi{\'c}}, {Barret}, {Bauer}, {Blandford}, {Brandt}, {Brenneman},
  {Chiang}, {Chakrabarty}, {Chenevez}, {Comastri}, {Dufour}, {Elvis}, {Fabian},
  {Farrah}, {Fryer}, {Gotthelf}, {Grindlay}, {Helfand}, {Krivonos}, {Meier},
  {Miller}, {Natalucci}, {Ogle}, {Ofek}, {Ptak}, {Reynolds}, {Rigby},
  {Tagliaferri}, {Thorsett}, {Treister}, \& {Urry}}]{harrison13}
{Harrison}, F.~A., {Craig}, W.~W., {Christensen}, F.~E., {et~al.} 2013, \apj,
  770, 103

\bibitem[{{Haslam} {et~al.}(1982){Haslam}, {Salter}, {Stoffel}, \&
  {Wilson}}]{haslamsky}
{Haslam}, C.~G.~T., {Salter}, C.~J., {Stoffel}, H., \& {Wilson}, W.~E. 1982,
  \aaps, 47, 1

\bibitem[{{Hobbs} {et~al.}(2020){Hobbs}, {Manchester}, {Dunning}, {Jameson},
  {Roberts}, {George}, {Green}, {Tuthill}, {Toomey}, {Kaczmarek}, {Mader},
  {Marquarding}, {Ahmed}, {Amy}, {Bailes}, {Beresford}, {Bhat}, {Bock},
  {Bourne}, {Bowen}, {Brothers}, {Cameron}, {Carretti}, {Carter}, {Castillo},
  {Chekkala}, {Cheng}, {Chung}, {Craig}, {Dai}, {Dawson}, {Dempsey}, {Doherty},
  {Dong}, {Edwards}, {Ergesh}, {Gao}, {Han}, {Hayman}, {Indermuehle},
  {Jeganathan}, {Johnston}, {Kanoniuk}, {Kesteven}, {Kramer}, {Leach},
  {Mcintyre}, {Moss}, {Os{\l}owski}, {Phillips}, {Pope}, {Preisig}, {Price},
  {Reeves}, {Reilly}, {Reynolds}, {Robishaw}, {Roush}, {Ruckley}, {Sadler},
  {Sarkissian}, {Severs}, {Shannon}, {Smart}, {Smith}, {Smith}, {Sobey},
  {Staveley-Smith}, {Tzioumis}, {van Straten}, {Wang}, {Wen}, \&
  {Whiting}}]{hobbs20}
{Hobbs}, G., {Manchester}, R.~N., {Dunning}, A., {et~al.} 2020, \pasa, 37, e012

\bibitem[{{Hu} {et~al.}(2020){Hu}, {Begi{\c{c}}arslan}, {G{\"u}ver}, {Enoto},
  {Younes}, {Sakamoto}, {Ray}, {Strohmayer}, {Guillot}, {Arzoumanian},
  {Palmer}, {Gendreau}, {Malacaria}, {Wadiasingh}, {Jaisawal}, \&
  {Majid}}]{hu20}
{Hu}, C.-P., {Begi{\c{c}}arslan}, B., {G{\"u}ver}, T., {et~al.} 2020, \apj,
  902, 1

\bibitem[{{Ibrahim} {et~al.}(2024){Ibrahim}, {Borghese}, {Coti Zelati},
  {Parent}, {Marino}, {Ould-Boukattine}, {Rea}, {Ascenzi}, {Pacholski},
  {Mereghetti}, {Israel}, {Tiengo}, {Possenti}, {Burgay}, {Turolla}, {Zane},
  {Esposito}, {G{\"o}tz}, {Campana}, {Kirsten}, {Gawro{\'n}ski}, \&
  {Hessels}}]{ibrahim24}
{Ibrahim}, A.~Y., {Borghese}, A., {Coti Zelati}, F., {et~al.} 2024, \apj, 965,
  87

\bibitem[{{Kaastra} \& {Bleeker}(2016)}]{kaastra16}
{Kaastra}, J.~S. \& {Bleeker}, J.~A.~M. 2016, \aap, 587, A151

\bibitem[{{Kaspi} \& {Beloborodov}(2017)}]{kaspi17}
{Kaspi}, V.~M. \& {Beloborodov}, A.~M. 2017, \araa, 55, 261

\bibitem[{{Klingler} {et~al.}(2021){Klingler}, {Page}, {Palmer}, \& {Neil
  Gehrels Swift Observatory Team}}]{klingler21}
{Klingler}, N.~J., {Page}, K.~L., {Palmer}, D.~M., \& {Neil Gehrels Swift
  Observatory Team}. 2021, GRB Coordinates Network, 30796, 1

\bibitem[{{Kovlakas} {et~al.}(2025){Kovlakas}, {De Grandis}, \&
  {Rea}}]{kovlakas2025}
{Kovlakas}, K., {De Grandis}, D., \& {Rea}, N. 2025, arXiv e-prints,
  arXiv:2509.03090

\bibitem[{{Kumar} {et~al.}(2021){Kumar}, {Shannon}, {Flynn}, {Os{\l}owski},
  {Bhandari}, {Day}, {Deller}, {Farah}, {Kaczmarek}, {Kerr}, {Phillips},
  {Price}, {Qiu}, \& {Thyagarajan}}]{kumar21}
{Kumar}, P., {Shannon}, R.~M., {Flynn}, C., {et~al.} 2021, \mnras, 500, 2525

\bibitem[{{Leahy} {et~al.}(1983){Leahy}, {Darbro}, {Elsner}, {Weisskopf},
  {Kahn}, {Sutherland}, \& {Grindlay}}]{leahy83}
{Leahy}, D.~A., {Darbro}, W., {Elsner}, R.~F., {et~al.} 1983, \apj, 266, 160

\bibitem[{{Mereghetti} {et~al.}(2020){Mereghetti}, {Savchenko}, {Ferrigno},
  {G{\"o}tz}, {Rigoselli}, {Tiengo}, {Bazzano}, {Bozzo}, {Coleiro},
  {Courvoisier}, {Doyle}, {Goldwurm}, {Hanlon}, {Jourdain}, {von Kienlin},
  {Lutovinov}, {Martin-Carrillo}, {Molkov}, {Natalucci}, {Onori}, {Panessa},
  {Rodi}, {Rodriguez}, {S{\'a}nchez-Fern{\'a}ndez}, {Sunyaev}, \&
  {Ubertini}}]{mereghetti20}
{Mereghetti}, S., {Savchenko}, V., {Ferrigno}, C., {et~al.} 2020, \apjl, 898,
  L29

\bibitem[{{Ocker} \& {Cordes}(2024)}]{ne2001p}
{Ocker}, S.~K. \& {Cordes}, J.~M. 2024, Research Notes of the American
  Astronomical Society, 8, 17

\bibitem[{{Palmer} {et~al.}(2021){Palmer}, {Evans}, {Kuin}, {Page}, \& {Swift
  Team}}]{palmer21a}
{Palmer}, D.~M., {Evans}, P.~A., {Kuin}, N.~P.~M., {Page}, K.~L., \& {Swift
  Team}. 2021, GRB Coordinates Network, 30120, 1

\bibitem[{{Pons} \& {Rea}(2012)}]{pons12}
{Pons}, J.~A. \& {Rea}, N. 2012, \apjl, 750, L6

\bibitem[{{Potekhin} {et~al.}(2015){Potekhin}, {Pons}, \&
  {Page}}]{potekhinponspage}
{Potekhin}, A.~Y., {Pons}, J.~A., \& {Page}, D. 2015, \ssr, 191, 239

\bibitem[{{Rajwade} {et~al.}(2022){Rajwade}, {Stappers}, {Lyne}, {Shaw},
  {Mickaliger}, {Liu}, {Kramer}, {Desvignes}, {Karuppusamy}, {Enoto},
  {G{\"u}ver}, {Hu}, \& {Surnis}}]{rajwade22}
{Rajwade}, K.~M., {Stappers}, B.~W., {Lyne}, A.~G., {et~al.} 2022, \mnras, 512,
  1687

\bibitem[{{Rea} {et~al.}(2020){Rea}, {Coti Zelati}, {Vigan{\`o}}, {Papitto},
  {Baganoff}, {Borghese}, {Campana}, {Esposito}, {Haggard}, {Israel},
  {Mereghetti}, {Mignani}, {Perna}, {Pons}, {Ponti}, {Stella}, {Torres},
  {Turolla}, \& {Zane}}]{rea20}
{Rea}, N., {Coti Zelati}, F., {Vigan{\`o}}, D., {et~al.} 2020, \apj, 894, 159

\bibitem[{{Rea} \& {De Grandis}(2025)}]{rea25}
{Rea}, N. \& {De Grandis}, D. 2025, arXiv e-prints, arXiv:2503.04442

\bibitem[{{Scholz} {et~al.}(2017){Scholz}, {Camilo}, {Sarkissian}, {Reynolds},
  {Levin}, {Bailes}, {Burgay}, {Johnston}, {Kramer}, \& {Possenti}}]{scholz17}
{Scholz}, P., {Camilo}, F., {Sarkissian}, J., {et~al.} 2017, \apj, 841, 126

\bibitem[{{Tiengo} {et~al.}(2010){Tiengo}, {Vianello}, {Esposito},
  {Mereghetti}, {Giuliani}, {Costantini}, {Israel}, {Stella}, {Turolla},
  {Zane}, {Rea}, {G{\"o}tz}, {Bernardini}, {Moretti}, {Romano}, {Ehle}, \&
  {Gehrels}}]{tiengo10}
{Tiengo}, A., {Vianello}, G., {Esposito}, P., {et~al.} 2010, \apj, 710, 227

\bibitem[{{Turolla} {et~al.}(2015){Turolla}, {Zane}, \& {Watts}}]{turolla15}
{Turolla}, R., {Zane}, S., \& {Watts}, A.~L. 2015, Reports on Progress in
  Physics, 78, 116901

\bibitem[{{Ubertini} {et~al.}(2003){Ubertini}, {Lebrun}, {Di Cocco}, {Bazzano},
  {Bird}, {Broenstad}, {Goldwurm}, {La Rosa}, {Labanti}, {Laurent}, {Mirabel},
  {Quadrini}, {Ramsey}, {Reglero}, {Sabau}, {Sacco}, {Staubert}, {Vigroux},
  {Weisskopf}, \& {Zdziarski}}]{ubertini03}
{Ubertini}, P., {Lebrun}, F., {Di Cocco}, G., {et~al.} 2003, \aap, 411, L131

\bibitem[{{Verner} {et~al.}(1996){Verner}, {Ferland}, {Korista}, \&
  {Yakovlev}}]{verner96}
{Verner}, D.~A., {Ferland}, G.~J., {Korista}, K.~T., \& {Yakovlev}, D.~G. 1996,
  \apj, 465, 487

\bibitem[{{Wilms} {et~al.}(2000){Wilms}, {Allen}, \& {McCray}}]{wilms00}
{Wilms}, J., {Allen}, A., \& {McCray}, R. 2000, \apj, 542, 914

\bibitem[{{Yakovlev} {et~al.}(2001){Yakovlev}, {Kaminker}, {Gnedin}, \&
  {Haensel}}]{yakovlev01}
{Yakovlev}, D.~G., {Kaminker}, A.~D., {Gnedin}, O.~Y., \& {Haensel}, P. 2001,
  \physrep, 354, 1

\bibitem[{{Younes} {et~al.}(2022){Younes}, {Lander}, {Baring}, {Enoto},
  {Kouveliotou}, {Wadiasingh}, {Ho}, {Harding}, {Arzoumanian}, {Gendreau},
  {G{\"u}ver}, {Hu}, {Malacaria}, {Ray}, \& {Strohmayer}}]{younes22}
{Younes}, G., {Lander}, S.~K., {Baring}, M.~G., {et~al.} 2022, \apjl, 924, L27

\bibitem[{{Zhang} {et~al.}(2020){Zhang}, {Li}, {Lu}, {Song}, {Xu}, {Liu},
  {Chen}, {Cao}, {Bu}, {Chang}, {Chen}, {Chen}, {Chen}, {Chen}, {Chen}, {Cui},
  {Cui}, {Deng}, {Dong}, {Du}, {Fu}, {Gao}, {Gao}, {Gao}, {Ge}, {Gu}, {Guan},
  {Gungor}, {Guo}, {Han}, {Hu}, {Huang}, {Huo}, {Jia}, {Jiang}, {Jiang}, {Jin},
  {Jin}, {Li}, {Li}, {Li}, {Li}, {Li}, {Li}, {Li}, {Li}, {Li}, {Li}, {Li},
  {Liang}, {Liao}, {Liu}, {Liu}, {Liu}, {Liu}, {Liu}, {Liu}, {Lu}, {Lu}, {Luo},
  {Ma}, {Meng}, {Nang}, {Nie}, {Ou}, {Qu}, {Sai}, {Shang}, {Shen}, {Sun},
  {Tan}, {Tao}, {Tuo}, {Wang}, {Wang}, {Wang}, {Wang}, {Wang}, {Wang}, {Wang},
  {Wen}, {Wu}, {Wu}, {Wu}, {Xiao}, {Xiong}, {Yan}, {Yang}, {Yang}, {Yang},
  {Yi}, {Yuan}, {Zhang}, {Zhang}, {Zhang}, {Zhang}, {Zhang}, {Zhang}, {Zhang},
  {Zhang}, {Zhang}, {Zhang}, {Zhang}, {Zhang}, {Zhang}, {Zhang}, {Zhang},
  {Zhang}, {Zhang}, {Zhang}, {Zhang}, {Zhang}, {Zhao}, {Zhao}, {Zheng}, {Zhou},
  {Zhu}, {Zhu}, {Zhuang}, \& {Insight-HXMT Team}}]{zhang20}
{Zhang}, S.-N., {Li}, T., {Lu}, F., {et~al.} 2020, Science China Physics,
  Mechanics, and Astronomy, 63, 249502

\bibitem[{{Zhang} {et~al.}(2022){Zhang}, {Xiao}, {Xiong}, {Cai}, {Zhang}, {Li},
  {Xie}, {Zhao}, {Huang}, {Song}, {Liu}, {Zhao}, {Guo}, {Zheng}, {Xue}, {Wang},
  {Yi}, {Zhang}, {Peng}, {Qiao}, {Guo}, {Li}, {Ma}, {Song}, {Wang}, {Wang},
  {Zhang}, {Zheng}, {Chen}, {He}, {Zhao}, {Du}, {Wu}, {Liang}, {Luo}, {Zhang},
  {Zhang}, {An}, {Gao}, {Gong}, {Li}, {Li}, {Li}, {Li}, {Li}, {Liang}, {Liu},
  {Liu}, {Sun}, {Tuo}, {Wang}, {Wen}, {Xu}, {Xu}, {Yang}, {Zhang}, {Zhang},
  {Zhang}, {Zhang}, {Zhou}, {Lu}, {Zhang}, \& {Gecam Team}}]{zhang2022_gcn}
{Zhang}, Y.~Q., {Xiao}, S., {Xiong}, S.~L., {et~al.} 2022, GRB Coordinates
  Network, 31397, 1

\bibitem[{{Zhang} {et~al.}(2021){Zhang}, {Xiong}, {Xiao}, {Cai}, {Zhao},
  {Huang}, {Song}, {Liu}, {Xie}, {Zhao}, {Guo}, {Zheng}, {Xue}, {Li}, {Wang},
  {Yi}, {Zhang}, {Zhang}, {Peng}, {Qiao}, {Guo}, {Li}, {Ma}, {Wang}, {Wang},
  {Zhang}, {Zheng}, {Chen}, {He}, {Zhao}, {Du}, {Wu}, {Liang}, {Luo}, {Zhang},
  {Song}, {Lu}, {Zhang}, \& {Gecam Team}}]{zhang21}
{Zhang}, Y.~Q., {Xiong}, S.~L., {Xiao}, S., {et~al.} 2021, GRB Coordinates
  Network, 30922, 1

\end{thebibliography}

\onecolumn
\begin{appendix}

\section{Log of X-ray and radio observations}

\footnotesize

\begin{ThreePartTable}

\begin{TableNotes}
\footnotesize
\item [a] The instrumental setup is indicated in brackets: PC = photon counting, WT = windowed timing.
\item [b] The count rate is in the 0.3--10\,keV energy range, except for \nicer\ (1.5--8\,keV for the first observation and 2--6\,keV for the remaining ones), \hxmt/LE (2.5--6.5\,keV), and \nustar\ (3--20\,keV). The upper limits are quoted at 3$\sigma$ c.l. and are derived from the stacked data.
\item [c] The flux is in the 0.3--10\,keV energy range. The upper limits for the archival observations are computed from the stack of the data assuming an absorbed blackbody spectrum with \nh $= 8.4\times10^{22}$\,cm$^{-2}$ and $kT_{\rm BB}=0.3$--0.5\,keV, and are quoted at 3$\sigma$ c.l..
\item [$^\dagger$] Data of these observations were merged for the spectral analysis.
\item [$^\ddagger$] The spectra of these observations were fitted jointly. For more detail, see Sec.\,\ref{sec:broadbandspec}.
\end{TableNotes}

\begin{longtable}{ccccccc}
\caption{X-ray observation log with observed and unabsorbed fluxes.}
\label{tab:obsX}\\
\scriptsize
X-ray Instrument$^a$ & Obs.ID & Start & Stop & Exposure & Net Count Rate$^b$ & Flux$^c$ (Obs / Unabs) \\  
  
                     &        &  \multicolumn{2}{c}{YYYY Mmm DD hh:mm:ss (TT)} & (ks) & (counts\,s$^{-1}$) & ($\times 10^{-11}$\,\flux) \\
\hline

\swift/XRT (PC) & 00042728001 & 2012 May 12 02:18:11 & 2012 May 12 02:27:55 & 0.6 & $<$0.007  & $<$0.03--0.04 / $<$0.9--0.2  \\
\swift/XRT (PC) & 00042729001 & 2012 May 13 05:35:25 & 2012 May 13 05:43:54 & 0.5 & $<$0.007  & $<$0.03--0.04 / $<$0.9--0.2   \\
\hline
\swift/XRT (PC) & 01053220000 & 2021 Jun 3 10:53:43 & 2021 Jun 3 11:21:26 & 1.7 & 0.30$\pm$0.01 & 3.9$\pm$0.3 / 6.2$\pm$0.3 \\
\nicer/XTI      & 4202190101 & 2021 Jun 3 11:21:31 & 2021 Jun 3 18:52:20 & 2.5 & 3.88$\pm$0.05 & 4.7$\pm$0.1 / 8.2$\pm$0.1 \\
\swift/XRT (WT) & 00014352001 & 2021 Jun 4 07:52:37 & 2021 Jun 4 09:36:56 & 2.0 & 0.42$\pm$0.02 & 4.7$\pm$0.3 / 7.9$\pm$0.3   \\
\swift/XRT (WT) & 00014352002 & 2021 Jun 5 10:42:46 & 2021 Jun 5 15:46:20 & 4.9 & 0.41$\pm$0.01 & 4.8$\pm$0.2 / 8.0$\pm$0.2 \\
\swift/XRT (WT) & 00014352003 & 2021 Jun 7 12:08:15 & 2021 Jun 7 13:48:55 & 1.9 & 0.42$\pm$0.02 & 4.9$\pm$0.3 / 8.2$\pm$0.3 \\
\hxmt/LE     & P0414008001 & 2021 Jun 10 17:39:42  & 2021 Jun 12 17:31:52  & 22.2 & 1.36$\pm$0.02 & -- \\
\swift/XRT (WT) & 00014352004 & 2021 Jun 11 11:46:41 & 2021 Jun 11 15:24:56 & 2.5 & 0.40$\pm$0.02 & 4.6$\pm$0.2 / 7.9$\pm$0.3 \\
\swift/XRT (WT) & 00014352005 & 2021 Jun 15 14:45:15 & 2021 Jun 15 16:34:55 & 1.6 & 0.41$\pm$0.02 & 4.8$\pm$0.3 / 8.0$\pm$0.4 \\
\swift/XRT (WT) & 00014352007 & 2021 Jun 21 15:17:01 & 2021 Jun 21 22:12:56 & 3.2 & 0.36$\pm$0.01 & 4.7$\pm$0.2 / 7.7$\pm$0.3 \\
\swift/XRT (PC) & 01057131000 & 2021 Jun 21 17:05:46 & 2021 Jun 21 17:30:08 & 1.5 & 0.32$\pm$0.02 & 3.7$\pm$0.2 / 5.8$\pm$0.3 \\
\swift/XRT (WT) & 00014352008 & 2021 Jun 24 16:38:23 & 2021 Jun 24 21:53:56 & 1.6 & 0.39$\pm$0.02 & 4.4$\pm$0.3 / 7.5$\pm$0.3 \\
\swift/XRT (WT) & 00014352009 & 2021 Jun 25 01:01:41 & 2021 Jun 25 11:46:56 & 1.7 & 0.38$\pm$0.02 & 4.3$\pm$0.3 / 7.3$\pm$0.3 \\
\swift/XRT (WT) & 00014352010 & 2021 Jun 26 14:43:25 & 2021 Jun 26 14:58:55 & 0.9 & 0.36$\pm$0.02 & 4.0$\pm$0.3 / 7.1$\pm$0.4 \\
\swift/XRT (WT) & 00014352011 & 2021 Jun 27 03:41:23 & 2021 Jun 27 13:16:56 & 2.1 & 0.38$\pm$0.02 & 4.2$\pm$0.2 / 7.1$\pm$0.3 \\
\swift/XRT (WT) & 00014352013 & 2021 Jun 30 01:31:24 & 2021 Jun 30 11:36:56 & 0.9 & 0.39$\pm$0.03 & 4.2$\pm$0.4 / 6.6$\pm$0.3 \\
\swift/XRT (WT) & 00014352014 & 2021 Jul 2 06:09:58 & 2021 Jul 2 09:37:56 & 2.7 & 0.35$\pm$0.01 & 4.2$\pm$0.2 / 6.8$\pm$0.3 \\
\swift/XRT (WT) & 00014352015 & 2021 Jul 4 02:40:47 & 2021 Jul 4 23:30:56 & 0.8 & 0.35$\pm$0.03 & 4.0$\pm$0.4 / 7.2$\pm$0.5 \\
\swift/XRT (WT) & 00014352016 & 2021 Jul 6 05:48:11 & 2021 Jul 6 06:07:56 & 1.2 & 0.34$\pm$0.02 & 4.1$\pm$0.3 / 6.9$\pm$0.4 \\
\swift/XRT (WT) & 00014352017 & 2021 Jul 8 13:48:13 & 2021 Jul 8 13:53:56 & 0.3 & 0.42$\pm$0.04 & 4.3$\pm$0.6 / 7.7$\pm$0.7 \\
\swift/XRT (WT) & 00014352018 & 2021 Jul 16 00:05:15  & 2021 Jul 17 03:15:56 & 2.1 & 0.35$\pm$0.02 & 3.9$\pm$0.2 / 6.8$\pm$0.3 \\
\swift/XRT (WT) & 00014352019 & 2021 Jul 20 06:01:59 & 2021 Jul 20 07:59:56  & 3.0 & 0.23$\pm$0.01 & 4.0$\pm$0.2 / 6.8$\pm$0.3 \\
\swift/XRT (WT) & 00014352020 & 2021 Jul 24 16:44:13 & 2021 Jul 24 16:52:56  & 0.5 & 0.45$\pm$0.04 & 5.6$\pm$0.6 / 8.6$\pm$0.7 \\
\swift/XRT (WT) & 00014352021 & 2021 Jul 26 21:24:21 & 2021 Jul 26 21:38:56  & 0.8 & 0.35$\pm$0.03 & 4.3$\pm$0.4 / 7.1$\pm$0.5 \\
\swift/XRT (WT) & 00014352022 & 2021 Jul 31 11:09:50 & 2021 Jul 31 20:46:56  & 0.4 & 0.25$\pm$0.03 & 3.2$\pm$0.5 / 5.6$\pm$0.6 \\
\swift/XRT (WT) & 00014352023 & 2021 Aug 1 22:12:39  & 2021 Aug 2 23:48:56   & 1.0 & 0.26$\pm$0.02 & 3.7$\pm$0.3 / 6.9$\pm$0.5 \\
\swift/XRT (WT) & 00014352024 & 2021 Aug 4 21:50:55  & 2021 Aug 4 21:55:56   & 0.3 & 0.11$\pm$0.04 & 2.0$\pm$0.6 / 3.9$\pm$0.7 \\
\swift/XRT (WT) & 00014352025 & 2021 Aug 7 01:00:00  & 2021 Aug 7 01:03:56   & 0.2 & 0.27$\pm$0.05 & 3.2$\pm$0.7 / 6.1$\pm$0.8 \\
\swift/XRT (WT) & 00014352026 & 2021 Aug 10 02:01:55 & 2021 Aug 10 10:09:56  & 1.1 & 0.32$\pm$0.02 & 4.3$\pm$0.4 / 6.8$\pm$0.4 \\
\swift/XRT (WT) & 00014352027 & 2021 Aug 11 00:29:41 & 2021 Aug 11 18:18:56  & 1.2 & 0.34$\pm$0.02 & 4.3$\pm$0.4 / 6.8$\pm$0.4 \\
\swift/XRT (WT) & 00014352028 & 2021 Aug 14 03:17:21 & 2021 Aug 14 23:59:56  & 1.9 & 0.33$\pm$0.02 & 3.6$\pm$0.2 / 6.2$\pm$0.3 \\
\swift/XRT (WT) & 00014352029 & 2021 Aug 23 11:50:03 & 2021 Aug 23 13:52:56  & 1.8 & 0.26$\pm$0.01 & 3.6$\pm$0.3 / 5.9$\pm$0.3 \\
\swift/XRT (PC) & 01070298000 & 2021 Aug 25 06:55:27 & 2021 Aug 25 08:18:42  & 1.7 & 0.29$\pm$0.01 & 3.4$\pm$0.2 / 5.3$\pm$0.3 \\
\swift/XRT (WT) & 00014352030 & 2021 Aug 28 03:22:22 & 2021 Aug 28 08:09:56  & 0.9 & 0.30$\pm$0.02 & 3.8$\pm$0.4 / 6.1$\pm$0.4 \\
\swift/XRT (WT) & 00014352031 & 2021 Aug 29 09:32:37 & 2021 Aug 30 03:13:56  & 1.7 & 0.31$\pm$0.02 & 3.6$\pm$0.2 / 6.2$\pm$0.3 \\
\swift/XRT (WT) & 00014352032 & 2021 Sep 3 09:10:14  & 2021 Sep 3 12:39:56   & 2.7 & 0.30$\pm$0.01 & 4.1$\pm$0.2 / 6.5$\pm$0.3 \\
\swift/XRT (PC) & 01072706000 & 2021 Sep 10 10:13:45 & 2021 Sep 10 10:34:51 & 1.3 & 0.20$\pm$0.01 & 3.7$\pm$0.4 / 5.9$\pm$0.4 \\
\swift/XRT (WT) & 00014352034 & 2021 Sep 13 05:02:19 & 2021 Sep 13 16:34:56 & 2.7 & 0.29$\pm$0.01 & 3.6$\pm$0.2 / 6.1$\pm$0.2 \\
\swift/XRT (WT) & 00014352035$^\dagger$ & 2021 Sep 17 17:19:51 & 2021 Sep 17 17:42:14 & 1.3 & 0.26$\pm$0.02 & 3.1$\pm$0.2 / 5.3$\pm$0.3 \\
\swift/XRT (WT) & 00014352036$^\dagger$ & 2021 Sep 22 01:03:52 & 2021 Sep 22 01:09:56 & 0.4 & 0.28$\pm$0.04 & 3.1$\pm$0.2 / 5.3$\pm$0.3 \\
\swift/XRT (WT) & 00014352037 & 2021 Sep 24 00:31:59 & 2021 Sep 24 13:47:56 & 2.8 & 0.19$\pm$0.01 & 3.1$\pm$0.2 / 5.2$\pm$0.3 \\
\swift/XRT (WT) & 00014352038 & 2021 Sep 30 09:37:51 & 2021 Sep 30 11:10:56 & 0.6 & 0.27$\pm$0.03 & 3.5$\pm$0.4 / 6.1$\pm$0.5 \\
\swift/XRT (WT) & 00014352039 & 2021 Oct 5 04:01:23 & 2021 Oct 5 21:54:56 & 2.4 & 0.27$\pm$0.01 & 3.4$\pm$0.2 / 5.4$\pm$0.3 \\
\swift/XRT (WT) & 00014352040$^\ddagger$ & 2021 Oct 7 08:36:19 & 2021 Oct 8 23:03:56 & 3.1 & 0.28$\pm$0.01 & 3.2$\pm$0.1 / 5.5$\pm$0.1 \\
\nustar/FPMA & 80702313006$^\ddagger$ & 2021 Oct 7 14:31:09 & 2021 Oct 8 19:16:09 & 51.6 & 0.706$\pm$0.004 & 3.2$\pm$0.1 / 5.5$\pm$0.1\\
\nustar/FPMB & 80702313006$^\ddagger$ & 2021 Oct 7 14:31:09 & 2021 Oct 8 19:16:09 & 51.5 & 0.640$\pm$0.004 & 3.2$\pm$0.1 / 5.5$\pm$0.1\\
\swift/XRT (WT) & 00014352041 & 2021 Oct 15 01:45:04 & 2021 Oct 15 22:25:56 & 2.7 & 0.26$\pm$0.01 & 3.1$\pm$0.2 / 5.2$\pm$0.2 \\
\swift/XRT (WT) & 00014352042 & 2021 Oct 22 01:01:02 & 2021 Oct 22 10:31:56 & 1.2 & 0.28$\pm$0.02 & 3.6$\pm$0.3 / 6.0$\pm$0.4 \\
\swift/XRT (PC) & 00014971001 & 2022 Jan 5 07:07:33 & 2022 Jan 5 21:42:52 & 3.2 & 0.18$\pm$0.01 & 2.2$\pm$0.1 / 3.5$\pm$0.2 \\
\swift/XRT (WT) & 00014971004 & 2022 Feb 19 11:22:27 & 2022 Feb 19 14:53:56 & 2.9 & 0.17$\pm$0.01 & 1.8$\pm$0.1 / 3.2$\pm$0.2 \\
\swift/XRT (WT) & 00014971005 & 2022 Mar 6 00:05:15 & 2022 Mar 6 19:35:56 & 2.2 & 0.13$\pm$0.01 & 1.9$\pm$0.2 / 3.2$\pm$0.2 \\
\swift/XRT (WT) & 00014971006 & 2022 Mar 21 15:38:22 & 2022 Mar 21 19:04:56 & 1.0 & 0.12$\pm$0.02 & 1.2$\pm$0.2 / 2.7$\pm$0.3 \\
\swift/XRT (WT) & 00014971007 & 2022 Apr 5 00:07:14 & 2022 Apr 5 17:45:56 & 2.4 & 0.13$\pm$0.01 & 1.9$\pm$0.2 / 3.0$\pm$0.2 \\
\swift/XRT (WT) & 00014971008 & 2022 Apr 20 14:15:53 & 2022 Apr 20 23:45:56 & 2.5 & 0.12$\pm$0.01 & 1.7$\pm$0.1 / 3.1$\pm$0.2 \\
\swift/XRT (WT) & 00014971009 & 2022 May 5 13:24:19 & 2022 May 5 15:37:56 & 2.7 & 0.12$\pm$0.01 & 1.7$\pm$0.1 / 2.8$\pm$0.2 \\
\swift/XRT (WT) & 00014971010 & 2022 May 20 10:23:49 & 2022 May 20 23:17:56 & 2.6 & 0.11$\pm$0.01 & 1.5$\pm$0.1 / 2.6$\pm$0.2 \\
\swift/XRT (WT) & 00014971011$^\dagger$ & 2022 Jun 4 05:15:15 & 2022 Jun 4 08:42:56 & 1.2 & 0.11$\pm$0.02 & 1.6$\pm$0.2 / 2.5$\pm$0.2 \\
\swift/XRT (WT) & 00014971012$^\dagger$ & 2022 Jun 6 17:52:38 & 2022 Jun 6 18:10:55 & 1.1 & 0.11$\pm$0.02 & 1.6$\pm$0.2 / 2.5$\pm$0.2 \\
\swift/XRT (WT) & 00014971013 & 2022 Jun 19 11:11:07 & 2022 Jun 19 13:04:56 & 2.6 & 0.14$\pm$0.01 & 1.5$\pm$0.2 / 2.4$\pm$0.2 \\
\swift/XRT (WT) & 00014971014 & 2022 Jul 4 02:54:17  & 2022 Jul 4 04:46:56  & 2.7 & 0.11$\pm$0.01 & 1.1$\pm$0.2 / 1.8$\pm$0.1 \\
\swift/XRT (WT) & 00014971015 & 2022 Jul 19 05:45:31 & 2022 Jul 19 09:19:56 & 2.9 & 0.08$\pm$0.01 & 0.9$\pm$0.1 / 1.8$\pm$0.1 \\
\nicer/XTI      & 5202190101$^\dagger$  & 2022 Jul 21 19:10:00 & 2022 Jul 21 22:29:20 & 0.9 & 0.85$\pm$0.03 & 1.3$\pm$0.1 / 2.2$\pm$0.1 \\
\nicer/XTI      & 5202190102$^\dagger$  & 2022 Jul 21 23:50:21 & 2022 Jul 22 03:07:20 & 0.7 & 0.96$\pm$0.04 & 1.3$\pm$0.1 / 2.2$\pm$0.1 \\
\nicer/XTI      & 5202190103 & 2022 Jul 27 07:38:20 & 2022 Jul 27 10:58:40 & 1.9 & 0.68$\pm$0.02 & 1.0$\pm$0.1 / 1.7$\pm$0.1 \\
\nicer/XTI      & 5202190104 & 2022 Jul 31 01:28:00 & 2022 Jul 31 03:25:40 & 1.2 & 0.79$\pm$0.11 & -- \\
\swift/XRT (WT) & 00014971016 & 2022 Aug 3 05:18:41 & 2022 Aug 3 15:05:56 & 2.9 & 0.07$\pm$0.01 & 0.9$\pm$0.1 / 1.6$\pm$0.1 \\
\nicer/XTI      & 5202190105  & 2022 Aug 5 16:12:00 & 2022 Aug 5 20:45:20 & 2.4 & 0.78$\pm$0.04 & 1.0$\pm$0.1 / 1.8$\pm$0.1 \\
\nicer/XTI      & 5202190106  & 2022 Aug 7 20:28:16 & 2022 Aug 7 23:53:20 & 1.4 & 1.46$\pm$0.04 & 2.3$\pm$0.1 / 3.7$\pm$0.1 \\
\nicer/XTI      & 5202190107$^\dagger$  & 2022 Aug 9 23:31:53 & 2022 Aug 9 23:41:06 & 0.4 & 0.85$\pm$0.05 & 1.1$\pm$0.1 / 1.8$\pm$0.1 \\
\nicer/XTI      & 5202190108$^\dagger$  & 2022 Aug 10 01:04:40 & 2022 Aug 10 01:14:06 & 0.4 & 0.80$\pm$0.05 & 1.1$\pm$0.1 / 1.8$\pm$0.1 \\
\nicer/XTI      & 5202190109  & 2022 Aug 11 11:17:20 & 2022 Aug 11 14:36:56 & 2.5 & 0.90$\pm$0.02 & 1.4$\pm$0.1 / 2.2$\pm$0.1 \\
\nicer/XTI      & 5202190110  & 2022 Aug 13 18:49:38 & 2022 Aug 13 19:04:58 & 0.6 & 0.80$\pm$0.04 & 1.2$\pm$0.1 / 1.9$\pm$0.1 \\
\nicer/XTI      & 5202190111$^\dagger$  & 2022 Aug 15 00:16:20 & 2022 Aug 15 00:51:40 & 0.9 & 0.64$\pm$0.04 & 1.2$\pm$0.1 / 1.8$\pm$0.1 \\
\nicer/XTI      & 5202190112$^\dagger$  & 2022 Aug 17 00:31:58 & 2022 Aug 17 02:26:38 & 0.6 & 0.80$\pm$0.05 & 1.2$\pm$0.1 / 1.8$\pm$0.1 \\
\swift/XRT (PC) & 00014971017 & 2022 Aug 18 00:15:37 & 2022 Aug 18 22:44:52 & 2.5 & 0.071$\pm$0.005 & 0.9$\pm$0.1 / 1.5$\pm$0.1 \\
\swift/XRT (PC) & 00014971018 & 2022 Sep 2 14:14:02  & 2022 Sep 2 16:13:54 & 2.7 & 0.067$\pm$0.005 & 0.9$\pm$0.1 / 1.3$\pm$0.1 \\
\swift/XRT (PC) & 00014971019 & 2022 Sep 16 17:25:36 & 2022 Sep 17 21:42:52 & 1.5 & 0.071$\pm$0.007 & 1.2$\pm$0.3 / 1.7$\pm$0.3 \\
\swift/XRT (PC) & 00014971020 & 2022 Sep 21 12:00:56  & 2022 Sep 21 13:42:52 & 1.1 & 0.042$\pm$0.006 & 0.7$\pm$0.2 / 1.1$\pm$0.2 \\
\swift/XRT (PC) & 00014971021 & 2022 Oct 2 21:20:51  & 2022 Oct 2 23:10:53 & 2.8 &  0.067$\pm$0.005 & 0.8$\pm$0.1 / 1.3$\pm$0.1 \\
\swift/XRT (PC) & 00014971022 & 2022 Oct 17 03:11:56 & 2022 Oct 17 09:52:51 & 3.2 & 0.053$\pm$0.004 & 1.1$\pm$0.2 / 1.5$\pm$0.2 \\
\swift/XRT (PC) & 00014971023 & 2023 Feb 8 05:27:27 & 2023 Feb 8 20:08:53 & 2.7 & 0.036$\pm$0.004 & 0.4$\pm$0.1 / 0.7$\pm$0.1 \\
\swift/XRT (PC) & 00014971024 & 2023 Feb 22 00:03:15 & 2023 Feb 23 23:29:52 & 2.7 & 0.035$\pm$0.004 & 0.5$\pm$0.1 / 0.7$\pm$0.1 \\
\swift/XRT (PC) & 00014971025 & 2023 Mar 10 11:21:24 & 2023 Mar 10 18:03:54 & 2.7 & 0.027$\pm$0.003 & 0.29$\pm$0.04 / 0.6$\pm$0.1 \\
\swift/XRT (PC) & 00014971026$^\dagger$ & 2023 Mar 25 04:15:22 & 2023 Mar 25 10:55:52 & 1.6 & 0.019$\pm$0.004 & 0.19$\pm$0.04 / 0.4$\pm$0.1 \\
\swift/XRT (PC) & 00014971027$^\dagger$ & 2023 Mar 29 20:55:49 & 2023 Mar 29 21:16:52 & 1.3 & 0.022$\pm$0.004 & 0.19$\pm$0.04 / 0.4$\pm$0.1 \\
\swift/XRT (PC) & 00014971028 & 2023 Apr 9 14:05:29 & 2023 Apr 9 20:43:53 & 2.7 & 0.020$\pm$0.003 & 0.3$\pm$0.1 / 0.4$\pm$0.1 \\
\swift/XRT (PC) & 00014971029 & 2023 Apr 24 00:27:15 & 2023 Apr 24 14:58:52 & 2.7 & 0.020$\pm$0.003 & 0.3$\pm$0.1 / 0.5$\pm$0.1 \\
\swift/XRT (PC) & 00014971030$^\dagger$ & 2023 May 09 06:48:03 & 2023 May 09 21:32:53 & 1.7 & 0.011$\pm$0.003 & 0.19$\pm$0.04 / 0.4$\pm$0.1 \\
\swift/XRT (PC) & 00014971031$^\dagger$ & 2023 May 24 03:03:39 & 2023 May 24 23:42:51 & 2.3 & 0.015$\pm$0.003 & 0.19$\pm$0.04 / 0.4$\pm$0.1 \\
\swift/XRT (PC) & 00014971032$^\dagger$ & 2023 Jun 08 10:33:35 & 2023 Jun 08 15:45:52 & 2.5 & 0.021$\pm$0.003 & 0.26$\pm$0.04 / 0.42$\pm$0.04 \\ 
\swift/XRT (PC) & 00014971033$^\dagger$ & 2023 Jun 23 10:40:47 & 2023 Jun 24 02:50:52 & 2.8 & 0.018$\pm$0.003 & 0.26$\pm$0.04 / 0.42$\pm$0.04 \\ 
\swift/XRT (PC) & 00014971034$^\dagger$ & 2023 Jul 08 07:57:13 & 2023 Jul 08 17:49:53 & 2.3 & 0.011$\pm$0.003 & 0.18$\pm$0.03 / 0.30$\pm$0.04 \\ 
\swift/XRT (PC) & 00014971035$^\dagger$ & 2023 Jul 23 05:07:40 & 2023 Jul 24 09:32:52 & 1.6 & 0.018$\pm$0.004 & 0.18$\pm$0.03 / 0.30$\pm$0.04 \\  
\swift/XRT (PC) & 00014971036$^\dagger$ & 2023 Jul 26 02:40:36 & 2023 Jul 26 18:50:52 & 1.8 & 0.016$\pm$0.003 & 0.18$\pm$0.03 / 0.30$\pm$0.04 \\  
\swift/XRT (PC) & 00014971037$^\dagger$ & 2023 Aug 07 03:23:20 & 2023 Aug 07 17:48:53 & 3.2 & 0.012$\pm$0.002 & 0.14$\pm$0.03 / 0.24$\pm$0.03 \\
\swift/XRT (PC) & 00014971038$^\dagger$ & 2023 Aug 22 12:51:21 & 2023 Aug 22 19:29:52 & 2.5 & 0.010$\pm$0.002 & 0.14$\pm$0.03 / 0.24$\pm$0.03 \\
\swift/XRT (PC) & 00014971039$^\dagger$ & 2023 Sep 6 01:57:21 & 2023 Sep 6 11:34:52 & 2.0 & 0.010$\pm$0.002 & 0.09$\pm$0.02 / 0.17$\pm$0.02 \\
\swift/XRT (PC) & 00014971040$^\dagger$ & 2023 Sep 21 04:51:56 & 2023 Sep 21 05:08:51 & 1.0 & 0.007$\pm$0.003 & 0.09$\pm$0.02 / 0.17$\pm$0.02 \\ 
\swift/XRT (PC) & 00014971041$^\dagger$ & 2023 Sep 26 06:42:47 & 2023 Sep 26 07:10:52 & 1.7 & 0.007$\pm$0.002 & 0.09$\pm$0.02 / 0.17$\pm$0.02 \\
\swift/XRT (PC) & 00014971042$^\dagger$ & 2023 Oct 6 05:57:57 & 2023 Oct 6 07:54:52 & 0.7 & 0.004$\pm$0.002 & 0.09$\pm$0.02 / 0.17$\pm$0.02 \\ 
\swift/XRT (PC) & 00014971043$^\dagger$ & 2023 Oct 12 00:09:16 & 2023 Oct 12 11:26:52 & 1.5 & 0.014$\pm$0.003 & 0.09$\pm$0.02 / 0.17$\pm$0.02 \\ 
\swift/XRT (PC) & 00014971044$^\dagger$ & 2023 Oct 21 04:22:16 & 2023 Oct 21 12:38:54 & 2.3 & 0.006$\pm$0.002 & 0.09$\pm$0.02 / 0.17$\pm$0.02 \\
\hline
\insertTableNotes  

\end{longtable}

\end{ThreePartTable}


\begin{sidewaystable*}
\caption{Radio observation log with limits on periodic and single-pulse radio emission}
\label{tab:obsradio} 
\centering
\begin{tabular}{ccccccc}
\hline        
\hline
Radio Instrument/backend & Frequency / Bandwidth & Start & Stop & Exposure & \multicolumn{2}{c}{Upper Limits\footnote{Upper limits (ULs) for the UWL are computed for the following sub-bands: 0.7-1.0\,GHz, 1.0-2.0\,GHz,2.0-4.0\,GHz. Both periodic emission and single pulse emission ULs are computed by exploiting the radiometer equation, assuming a S/N of 7 for pulsations and 6 for single pulses (see Sec.\,\ref{sec:radio}). We assumed a duty cycle of 10\% for the pulsar emission and 1\,ms for single pulses. System  and Sky temperature, for each band, are evaluated from \cite{hobbs20} and \cite{haslamsky}, respectively. To compensate for eventual pulse smearing, due to propagation effects, we estimated the dispersion and scattering by using the NE2001 model \citep{ne2001,ne2001p}.}} \\ 
        &          & \multicolumn{2}{c}{YYYY Mmm DD hh:mm:ss (TT)} & (hr) & Periodic Emission ($\mu$Jy) & Single Pulse (Jy\,ms) \\ 
\hline 
\pks/Medusa 	& 2.4\,GHz / 3.3\,GHz     & 2021 Jun 4 08:09        & 2021 Jun 4 10:18      & 2.16  & (163,34,19) & (1.6, 0.5, 0.2)\\  
\pks/PDFB4 	    & 1369\,MHz / 256\,MHz    & 2021 Jun 4 08:09        & 2021 Jun 4 10:18      & 2.16  & 140                                          & 1 \\   
\pks/Medusa 	& 2.4\,GHz / 3.3\,GHz     & 2021 Jun 5  10:18       & 2021 Jun 5 13:27      & 3.16  & (130, 27, 15) &  (1.6,0.5,0.2)\\  
\pks/PDFB4 	    & 3100\,MHz / 1024\,MHz   & 2021 Jun 5  10:18       & 2021 Jun 5 13:27      & 3.16  & 40                                        &  0.2  \\ 
\pks/Medusa     & 2.4\,GHz / 3.3\,GHz     & 2021 Jun 7 11:43        & 2021 Jun 7 14:57      & 3.24  & (130, 27, 15) &  (1.6,0.5,0.2)\\
\pks/PDFB4      & 1369\,MHz / 256\,MHz    & 2021 Jun 7 11:43        & 2021 Jun 7 14:57      & 3.24  & 110                                         & 1\\
\hline
\end{tabular}
\end{sidewaystable*}


\clearpage

\section{Journal of the X-ray bursts} 
Table\,\ref{tab:bursts} reports the times of arrival (in Barycentric Dynamical Time) and the fluence (in units of net counts) for all the X-ray bursts detected in the data presented in this work. The fluence values refer to the energy range 0.3--10\,keV. For the bursts detected by \swift/XRT, the fluence values may not reflect the true intrinsic fluence, due to uncertainties related to the detector saturation limits.

The duration of each burst was estimated either by summing the time bins at the finer resolution showing enhanced emission or by setting it equal to the coarser time resolution at which the burst is detected. Therefore, it has to be considered as an approximate value. Except for burst \#9 on 2021 June 5 (31.25\,ms), burst \#2 on June 21 (31.25\,ms), and burst \#1 on July 2 (125\,ms), burst \#1 on July 20 (39.06\,ms), bursts \#1 and \#3 on 2022 June 19 (125\,ms), all bursts have a duration of 62.5\,ms according to our definition.


\begin{longtable}{cccc}
\caption{Log of X-ray Bursts.}
\label{tab:bursts}\\
\footnotesize
Instrument & \multicolumn{2}{c}{Burst epoch} & Fluence \\  
           & \multicolumn{2}{c}{YYYY Mmm DD hh:mm:ss (TDB)} & (net counts) \\ 
\hline
\nicer/XTI      	& 2021 Jun 3 	& 13:59:21 	& 12  \\
                		& 			& 14:52:32 	& 26  \\ \hline
\swift/XRT (WT) 	& 2021 Jun 4	&  8:06:58 	& 6  \\
                	& 			& 9:31:13     	& 8  \\ 
                	& 			& 9:40:21     	& 9  \\ \hline
\swift/XRT (WT) & 2021 Jun 5  & 2021-06-05 11:02:14 & 8 \\
                &             & 11:11:10 & 10  \\
                &             & 11:12:17 & 7  \\
                &             & 11:15:50 & 6  \\
                &             & 12:24:35 & 6  \\
                &             & 12:29:03 & 6  \\
                &             & 14:04:04 & 12 \\
                &             & 15:42:27 & 8  \\
                &             & 15:46:41 & 20 \\
                &             & 15:49:07 & 7 \\ 	 \hline
\swift/XRT (WT) 	& 2021 Jun 7 	& 12:41:50 	& 6  \\
                	& 			& 12:42:35 	& 6  \\ \hline         
\swift/XRT (WT) 	& 2021 Jun 11  & 12:01:11 	& 8  \\
                	& 			& 12:06:46 	&  8  \\
                	& 			& 15:27:59 	& 6  \\ \hline
\swift/XRT (WT) 	& 2021 Jun 15  & 15:04:06 	& 6  \\ \hline
\swift/XRT (WT) 	& 2021 Jun 21  & 15:25:44 	& 10  \\
                    & 			& 15:36:38 	& 14 \\
                    & 			& 15:43:34 	& 6  \\
                    & 			& 15.43:35 	& 8  \\
                    & 			& 17.09:39 	& 7  \\ \hline
\swift/XRT (WT) 	& 2021 Jun 24 & 16:48:33 & 6 \\
                    &              & 16:52:46 & 9 \\
                    &              & 16:54:21 & 6 \\
                    &              & 16:56:33 & 6 \\ \hline
\swift/XRT (WT)     & 2021 Jun 25  & 10:33:48 & 6 \\                
                    &              & 10:36:00 & 6 \\ \hline
\swift/XRT (WT)     & 2021 Jun 27  & 03:51:57 & 16 \\
                    &              & 03:54:30 & 8 \\
                    &              & 03:54:55 & 7 \\
                    &              & 03:54:58 & 13 \\
                    &              & 11:41:23 & 9 \\ \hline
\swift/XRT (WT)     & 2021 Jun 30  & 01:50:37 & 56 \\ \hline
\swift/XRT (WT)     & 2021 Jul 2   & 06:35:06 & 17 \\ \hline
\swift/XRT (WT)     & 2021 Jul 20  & 07:57:05 &  21 \\ \hline
\swift/XRT (WT)     & 2021 Aug 7   & 01:07:16 & 8 \\ \hline
\swift/XRT (WT)     & 2021 Aug 10  & 02:06:38 & 9 \\ \hline
\swift/XRT (WT)     & 2021 Aug 29  & 09:45:11 & 11 \\
                    &              & 09:52:10 & 6 \\  \hline 
\swift/XRT (WT)     & 2021 Sep 3  & 12:39:32 & 11 \\ \hline
\swift/XRT (WT)     & 2021 Sep 24 & 13:31:09 & 17 \\ \hline
\swift/XRT(WT)      & 2022 Feb 19  & 11:31:07 & 15 \\
                    &              & 13:12:26 & 9 \\ \hline
\swift/XRT(WT)      & 2022 Mar 6   & 13:07:26 & 10 \\ \hline
\swift/XRT(WT)      & 2022 Apr 5   & 14:16:16 & 6 \\
                    &              & 14:20:51 & 6 \\ \hline
\swift/XRT(WT)      & 2022 May 5   & 15:32:54 & 9 \\ \hline         
\swift/XRT(WT)      & 2022 May 20  & 10:32:21 & 11 \\ \hline
\swift/XRT(WT)      & 2022 Jun 19  & 11:23:55 & 23 \\
                    &              & 11:27:31 & 19 \\
                    &              & 11:33:26 & 28 \\ \hline
\nicer/XTI          & 2022 Jul 21  & 13:59:21 & 12 \\
                    &              & 14:52:32 & 26 \\ \hline
\swift/XRT(WT)      & 2022 Aug 3   & 5:42:28 & 13 \\
                    &              & 11:44:37 & 16 \\
                    &              & 11:49:26 & 11 \\ 
\hline
\end{longtable}

\end{appendix}

\end{document}